\documentclass{aa}  
\pdfoutput=1

\usepackage{graphicx}
\usepackage{lscape}
\usepackage{rotating}
\usepackage[labelfont=bf]{caption}
\usepackage{subfig}
\usepackage{amsmath}
\usepackage{natbib}
\usepackage{txfonts}

\usepackage{hyperref}
\hypersetup{colorlinks=true,citecolor=blue,linkcolor=black}

\graphicspath{{figures/}}
\newcommand{\change}[1]{#1}

\begin{document}

   \title{Increased H$_2$CO production in the outer disk around HD 163296}


   \author{M.T. Carney \inst{1}, M.R. Hogerheijde \inst{1}, R.A. Loomis \inst{2},  \change{V.N. Salinas \inst{1},
   K.I. \"{O}berg \inst{2},  C. Qi \inst{2}, D.J. Wilner \inst{2}}
          }

   \institute{Leiden Observatory, Leiden University, PO Box 9513, 2300 RA, The Netherlands. \\
              \email{masoncarney@strw.leidenuniv.nl}
         \and
             Department of Astronomy, Harvard University, Cambridge, MA 02138, USA
             }

   \date{Received July 19, 2016 / Accepted May 25, 2017}

 
  \abstract
   {The gas and dust in circumstellar disks provide the raw materials to form planets. 
   The study of organic molecules and their building blocks in such disks offers insight 
   into the origin of the prebiotic environment of terrestrial planets.}
   {We aim to determine the distribution of formaldehyde, H$_2$CO, in the disk around HD 163296 
   to assess \change{the contribution of gas- and solid-phase formation routes of this simple} organic.}
   {\change{Three formaldehyde lines were observed (H$_2$CO 3$_{03}$--2$_{02}$, H$_2$CO 3$_{22}$--2$_{21}$, 
   and H$_2$CO 3$_{21}$--2$_{20}$) in the protoplanetary disk around the Herbig Ae star HD 163296 with ALMA
   at $\sim$0.5$\arcsec$ (60 AU) spatial resolution.
   Different parameterizations of the H$_2$CO abundance were compared to the observed visibilities, 
   using either a characteristic temperature, a characteristic radius or a radial power law index to 
   describe the H$_2$CO chemistry. Similar models were applied to ALMA Science Verification data of C$^{18}$O. 
   $\chi^2$ minimization on the visibilities was used to determine the best-fit model in each scenario.}
   }
   {\change{H$_2$CO 3$_{03}$--2$_{02}$ was readily detected via imaging, while the weaker
   H$_2$CO 3$_{22}$--2$_{21}$ and H$_2$CO 3$_{21}$--2$_{20}$ lines required matched filter analysis to detect.
   H$_2$CO is present throughout most of the gaseous disk, extending out to $\sim$550 AU.
   An apparent 50 AU inner radius of the H$_2$CO emission is likely caused by an optically thick dust continuum.
   The H$_2$CO radial intensity profile shows a peak at $\sim$100 AU and a secondary bump at $\sim$300 AU,
   suggesting increased production in the outer disk.
   In all modeling scenarios, fits to the H$_2$CO data show an increased abundance in the outer disk.
   The overall best-fit H$_2$CO model shows a factor of two enhancement
   beyond a radius of 270$\pm$20 AU, with an inner abundance (relative to H$_2$) of $2\!-\!5 \times 10^{-12}$.
   The H$_2$CO emitting region has a lower limit on the kinetic temperature of $T > 20$ K. 
   The C$^{18}$O modeling suggests an order of magnitude depletion of C$^{18}$O in the outer disk
   and an abundance of $4\!-\!12 \times 10^{-8}$ in the inner disk.}
   }
   {\change{There is a desorption front seen in the H$_2$CO emission that roughly coincides with the 
   outer edge of the 1.3 millimeter continuum. The increase in 
   H$_2$CO outer disk emission could be a result of hydrogenation
   of CO ices on dust grains that are then sublimated via thermal desorption or UV photodesorption, or 
   more efficient gas-phase production beyond $\sim$300 AU if CO is photodisocciated in this region.}
   }

   \authorrunning{Carney, M.~T. et al.}
   \titlerunning{H$_2$CO in HD 163296}
   
   \maketitle
%

\section{Introduction}

Protoplanetary disks have a layered temperature and density
structure that results in a cold, dense midplane where
gaseous molecules freeze out onto icy mantles around small
dust grains. Chemical reactions and radiative processing of 
atoms and molecules locked up in ices 
can create organic molecules of increasing complexity
\citep{Watanabe2003,Oberg2009,Oberg2010a,Herbst2009}.
The high densities and the vertical settling 
of larger grains make the disk midplane an ideal site for
grain growth and the formation of comets and planetesimals 
\citep{Dullemond2005,Andrews2005,DAlessio2006}.
The cold, complex molecular reservoir may be incorporated into
small icy bodies in the midplane and remain relatively
unprocessed, thus comets may preserve the chemical
composition of the disk at the time of their formation 
\citep{vanDishoeck1998,Mumma2011}.
Comets and other planetesimals are possible delivery mechanisms of organics
to terrestrial bodies during the early stages of
the solar system, thus it is important to understand
the chemistry and composition of their natal environments.
Observations of molecular emission lines can
determine the distribution and abundance of a molecular
species and constrain its location in a protoplanetary disk.
Characterizing simple organic molecules that
may be produced in the disk midplane, such as H$_2$CO,
can constrain available formation scenarios
for complex organic molecules (COMs). 
H$_2$CO acts as a precursor
to CH$_3$OH, which is an important building block
for other COMs \citep{Oberg2009,Walsh2014}. Thus, 
determining the dominant formation mechanism for H$_2$CO
and its distribution in disks can help to constrain 
abundances for CH$_3$OH and the 
complex organic reservoir.

\begin{table*}[!t]
 \caption{HD 163296 Observational Parameters}
 \centering
 \label{tab:obs_par}
 \begin{tabular}{lccc}
 \hline \hline
 \multicolumn{4}{c}{Project 2013.1.01268.S} \\
 \hline
 Dates Observed & \multicolumn{3}{c}{2014 July 27, 28, 29} \\
 Baselines & \multicolumn{3}{c}{21 -- 795 m | 16 -- 598 k${\rm \lambda}$ } \\
 \hline
  & H$_2$CO 3$_{03}$--2$_{02}$ & H$_2$CO 3$_{22}$--2$_{21}$ & H$_2$CO 3$_{21}$--2$_{20}$ \\
 Rest frequency [GHz] & 218.222 & 218.476 & 218.760  \\
 Synthesized beam [FWHM] & $0.54\arcsec \times 0.42\arcsec$ & $0.54\arcsec \times 0.42\arcsec$ & $0.53\arcsec \times0.42 \arcsec$ \\
 Position angle & 89.3$^\circ$ & 86.6$^\circ$ & 87.9$^\circ$ \\
 Channel width [km s$^{-1}$] & 0.084 & 0.084 & 0.084 \\
 rms noise\tablefootmark{a} [mJy beam$^{-1}$]  & 1.8 & 2.6 & 2.6 \\
 Integrated flux [Jy km s$^{-1}$]  & 0.64$\pm$0.06\tablefootmark{b}  & $>$0.036,$<$0.27\tablefootmark{c} & $>$0.032,$<$0.31\tablefootmark{c} \\
 Weighting & natural & natural & natural \\
 \hline
 Continuum Frequency [GHz] & \multicolumn{3}{c}{225.0}  \\
 Synthesized beam [FWHM] &  \multicolumn{3}{c}{$0.42\arcsec \times 0.33\arcsec$}  \\
 Position angle &  \multicolumn{3}{c}{77.5$^\circ$} \\
 rms noise [mJy beam$^{-1}$]  & \multicolumn{3}{c}{0.05} \\
 Integrated flux [mJy]  & \multicolumn{3}{c}{652$\pm$65}  \\
 Weighting & \multicolumn{3}{c}{Briggs, robust = 0.5} \\
 \hline \hline
 \multicolumn{4}{c}{Project 2011.1.00010.SV} \\
 \hline
 Dates Observed & \multicolumn{3}{c}{2012 June 09, 23, July 07} \\
 Baselines & \multicolumn{3}{c}{21 -- 536 m | 16 -- 402 k${\rm \lambda}$ } \\
 \hline 
  &  & C$^{18}$O 2--1 & \\
 Rest frequency [GHz] &  & 219.560 & \\
 Synthesized beam [FWHM] &  & $0.87\arcsec \times 0.71\arcsec$ & \\
 Position angle & & 64.0$^\circ$ &  \\
 Channel width [km s$^{-1}$] & & 0.334 &  \\
 rms noise\tablefootmark{a} [mJy beam$^{-1}$]  &  & 4.2 &  \\
 Integrated flux\tablefootmark{d} [Jy km s$^{-1}$]  & & 7.4$\pm$0.7 &  \\
 Weighting & & natural &  \\
 \hline
 \end{tabular}
 \tablefoot{Flux errors are dominated by systematic uncertainties, taken to be $\sim$10\%. 
 \tablefoottext{a}{Noise values are per image channel.}
 \tablefoottext{b}{\change{Line flux derived from spatial and spectral integration after 
 masking pixels with <3$\sigma$ emission.}}
 \tablefoottext{c}{Line flux lower limit derived from the peak $\sigma$-ratio based on matched-filter
 detections. Upper limits are 3$\sigma_{\rm I}$ where $\sigma_{\rm I}$ is defined in Section~\ref{sec:res_distr}.}
 \tablefoottext{d}{Line flux derived from spatial and spectral integration over 
 a 5.6$\arcsec$ aperture and velocity channels 0.87 km s$^{-1}$ -- 12.1 km s$^{-1}$.}
 }
\end{table*}

A major formation pathway of H$_2$CO is expected
to be the hydrogenation of CO ices in the cold midplane of the disk
\citep{Watanabe2003,Cuppen2009}. 
H$_2$CO also has a gas-phase formation route via neutral-neutral 
reactions of CH$_3$ and O at higher ($\gtrsim$200 K) 
temperatures \citep{Fockenberg2002,Atkinson2006}.
Formaldehyde has already been detected toward several protoplanetary disks
\citep[][\"{O}berg et al. in press]{Aikawa2003,Oberg2010b,Qi2013,vanderMarel2014,Loomis2015}, but
it is difficult to determine the contribution of H$_2$CO 
formed in the gas phase versus that formed via surface reactions.
It is important to consider the distribution of H$_2$CO in relation
to the freeze-out of CO, i.e. the CO snow line. H$_2$CO that exists
well beyond the CO snow line is likely formed on the icy mantles
of dust grains while H$_2$CO located within the CO snow line
forms via gas-phase pathways at higher temperatures.

\citet{Qi2013} attempted to reproduce Submillimeter Array (SMA)
observations of H$_2$CO around TW Hya
and HD 163296 with two simple parameterized models: a power-law
H$_2$CO column density with an inner radius and a 
ring-like H$_2$CO distribution with an upper boundary set by 
the CO freeze-out temperature.
They found that both models indicated H$_2$CO is produced mostly
at larger radii beyond the CO snow line in the disk around HD 163296,
consistent with a scenario where formaldehyde forms in CO ice and is
subsequently released back into the gas phase.
\citet{Loomis2015} modeled H$_2$CO in DM Tau observed with the 
Atacama Large Millimeter/submillimeter Array (ALMA) using a small
chemical network with and without grain-surface formation.
They found that both gas- and solid-phase production of H$_2$CO
were needed to reproduce the centrally peaked emission and the 
emission exterior to the CO snow line in DM Tau.

HD 163296 \change{(MWC 275)} is an ideal testbed for chemical processing in
protoplanetary disks, in particular for organics. It is
an isolated Herbig Ae protostar with spectral type A2Ve, an age of
approximately 5 Myr, located at 122 pc
\citep{deGregorioMonsalvo2013}. The protostar is surrounded
by a large gas-rich protoplanetary disk that extends to
$\sim$550 AU \citep{Isella2007} with stellar mass $M_*$ = 2.3  
$M_{\odot}$, disk mass $M_{\rm disk}$ = 0.089 $M_{\odot}$,
and an inclination of 44$^{\circ}$ based on the \citet{Qi2011} physical model. 
At such an inclination, vertical structure as well as radial structure
can be inferred from molecular line emission. The proximity
and size of the disk combined with the strong UV field of the
Herbig Ae protostar
provides a unique opportunity to fully resolve the
location of the CO snow line around HD 163296.
Several attempts have already been made to constrain the 
location of the CO snow line in this disk \citep{Qi2011,Mathews2013,Qi2015}.
Current estimates by \citet{Qi2015} place CO freeze-out at 90 AU, 
corresponding to $\sim$24 K in this disk.
HD 163296 is one of the best candidates to probe the formation of
organics with respect to the freeze-out of abundant volatiles 
like CO. Observations of H$_2$CO in combination with tracers of
the CO snow line, such as the optically thin C$^{18}$O isotopologue,
DCO$^+$, or N$_2$H$^+$,
provide insight into the formation of organic molecules
in Herbig Ae/Be disks.

This paper presents ALMA observations of H$_2$CO 
toward HD 163296 and characterizes its distribution
throughout the disk. Our analysis also makes use of C$^{18}$O 
Science Verification data, which has been previously reported
\citep{Rosenfeld2013,Qi2015}. Section~\ref{sec:obs} describes the observations
and data reduction. The detection of H$_2$CO, the modeling of H$_2$CO 
and C$^{18}$O distributions and abundances,
and the calculation of excitation temperatures for H$_2$CO are 
discussed in Section~\ref{sec:res}. In Section~\ref{sec:disc}
we discuss the relationship between H$_2$CO, C$^{18}$O, and the 
millimeter continuum, and the implications for H$_2$CO formation.


\section{Observations and Reduction}
\label{sec:obs}

HD 163296 (J2000: R.A. = 17$^{\rm{h}}$56$^{\rm{m}}$21.280$^{\rm{s}}$, 
DEC = --21$^\circ$57$\arcmin$22.441$\arcsec$) was observed on 2014 July 
27, 28, and 29 with ALMA in band 6 as part of Cycle 2. In total 33 antennas were used in 
the C34 configuration to achieve a resolution of $\sim$0.4$\arcsec$. 
Band 6 operates in the 211--275 GHz range as a 2SB receiver. 
The upper sideband contained continuum observations in the Time
Domain Mode (TDM) correlator setting with 128 channels over a 2 GHz 
bandwidth centered at 233 GHz, \change{presented in \citet{Zhang2016}.}
Three transitions of H$_2$CO were observed 
in the lower sideband with the Frequency Domain Mode (FDM) correlator setting:
H$_2$CO 3$_{03}$--2$_{02}$ at 218.22219 GHz, H$_2$CO 3$_{22}$--2$_{21}$ 
at 218.475632 GHz, and H$_2$CO 3$_{21}$--2$_{20}$ at 218.760066 GHz. 
Each line had a bandwidth of 56.6 MHz with 960 channels, 
providing a frequency (velocity) resolution of 0.061 MHz (0.084 km s$^{-1}$). 
Table~\ref{tab:obs_par} summarizes the observational parameters of each line. 
\change{Three additional lines, DCO$^+$ 3--2 at 216.11258 GHz, 
DCN 3--2 at 217.23853 GHz, and N$_2$D$^+$ 3--2 231.321828 GHz were also
observed with the same spectral parameters 
and will be presented in Salinas et al. (in prep).}

Visibility data were obtained over four execution blocks of 
$\sim$30 minutes (x1) and $\sim$90 minutes (x3) at 6.05 seconds 
per integration for 155 minutes total time on source. System temperatures
varied from 50--150 K. The average precipitable water vapor
across all observations was 1.0 mm. The Common Astronomy 
Software Applications (\textsc{casa}) package was used to 
calibrate the data with an automated script provided by the ALMA staff.
Calibration of each execution block was done with J1700-2610 as 
the delay calibrator, J1733-1304 as the bandpass and gain calibrator, 
J1733-1304 as the flux calibrator for three out of four blocks, 
and Titan as the flux calibrator for the final block. After initial 
calibration of individual execution blocks, gain calibration 
solutions obtained from models of Titan were used to derive fluxes 
for J1733-1304, which was then used as the flux calibrator 
in all spectral windows and all execution blocks for consistency. 
Amplitudes for HD 163296 were rescaled
across all blocks using J1733-1304 as the flux calibrator.
The average flux values for J1733-1304 were 
1.329 Jy in the lower sideband and 1.255 Jy in 
the upper sideband.  
The total flux for HD 163296 was found to be within 5\% across 
all execution blocks. All measurement sets were subsequently 
concatenated and time binned to 30s integration time per visibility 
for imaging and analysis.

\begin{figure}[!htbp]
 \centering
 \includegraphics[width=0.48\textwidth]{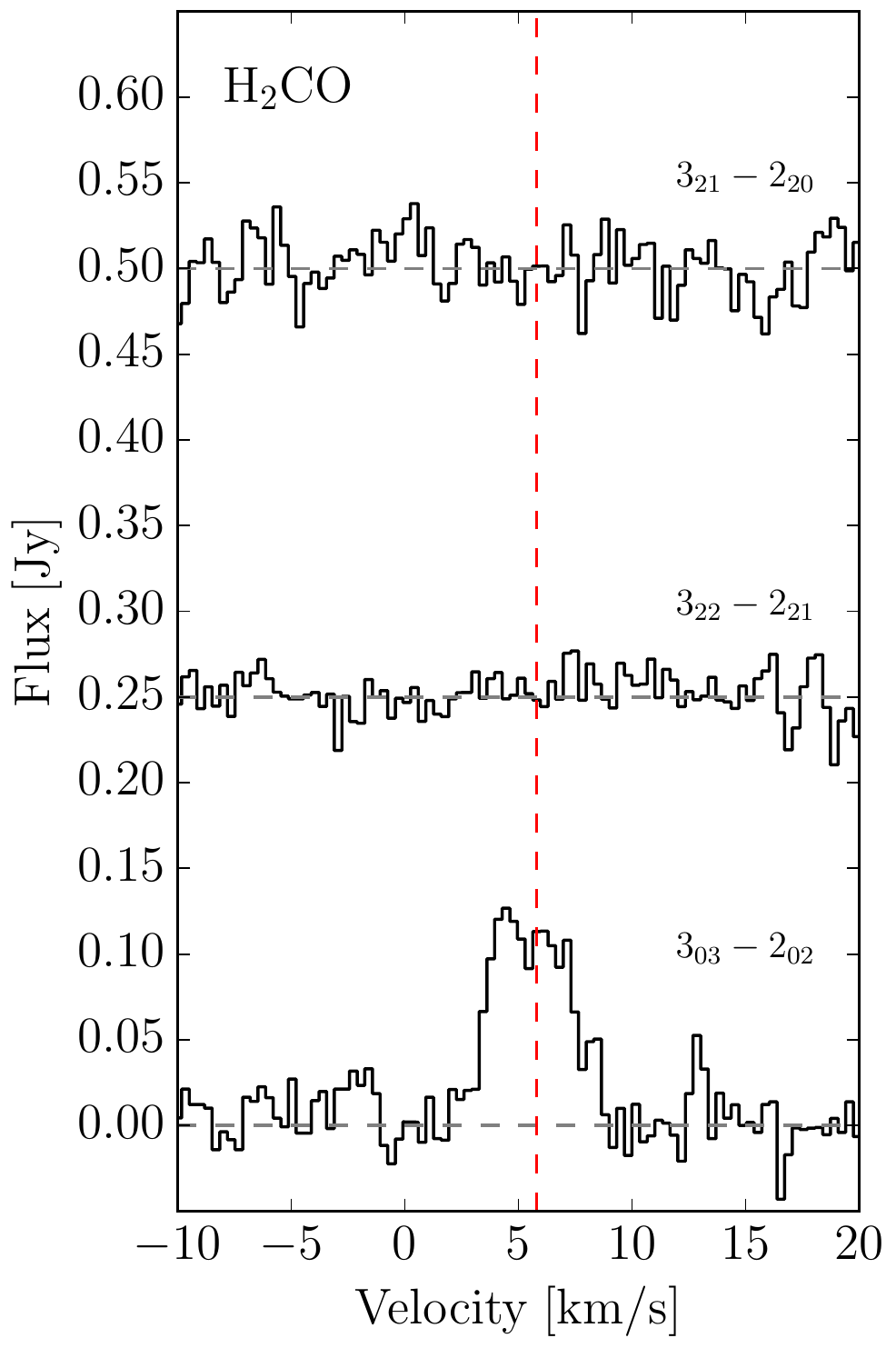}
 \caption{Disk-integrated H$_2$CO spectra \change{using a 5.6$\arcsec$ circular aperture}.
 H$_2$CO 3$_{03}$--2$_{02}$, H$_2$CO 3$_{22}$--2$_{21}$, 
 and H$_2$CO 3$_{21}$--2$_{20}$ are at y-offsets of 0, 0.25, and 0.5 Jy, respectively,
 \change{shown in dashed gray lines}. 
 The vertical dashed red line shows the systemic velocity. The spectra are Hanning smoothed
 to 0.336 km s$^{-1}$ channels.}
 \label{fig:spectra}
\end{figure}

\begin{figure*}[!htbp]
 \centering
 \includegraphics[width=0.3\textwidth]{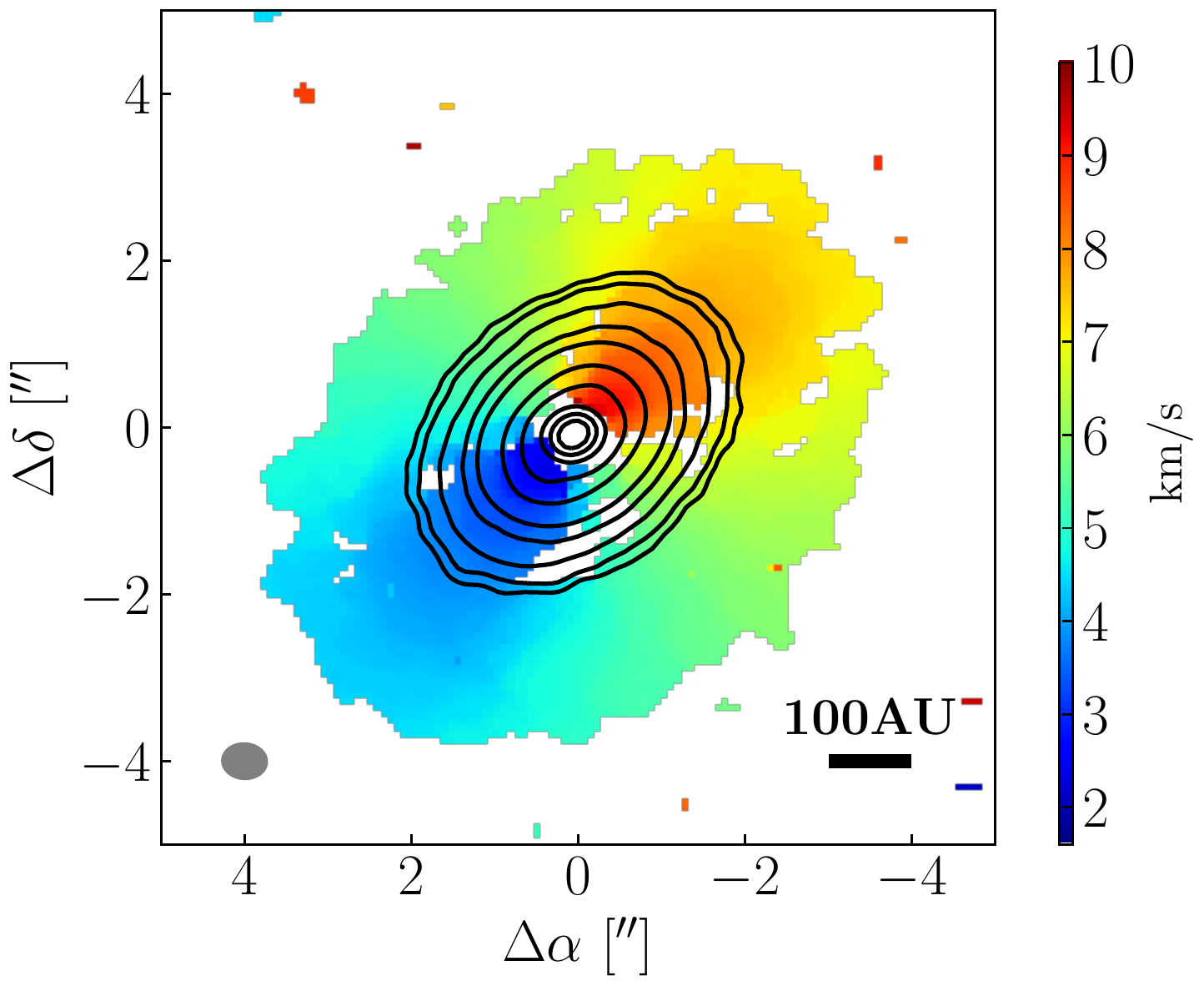}
 \hspace{0.1cm}
 \includegraphics[width=0.32\textwidth]{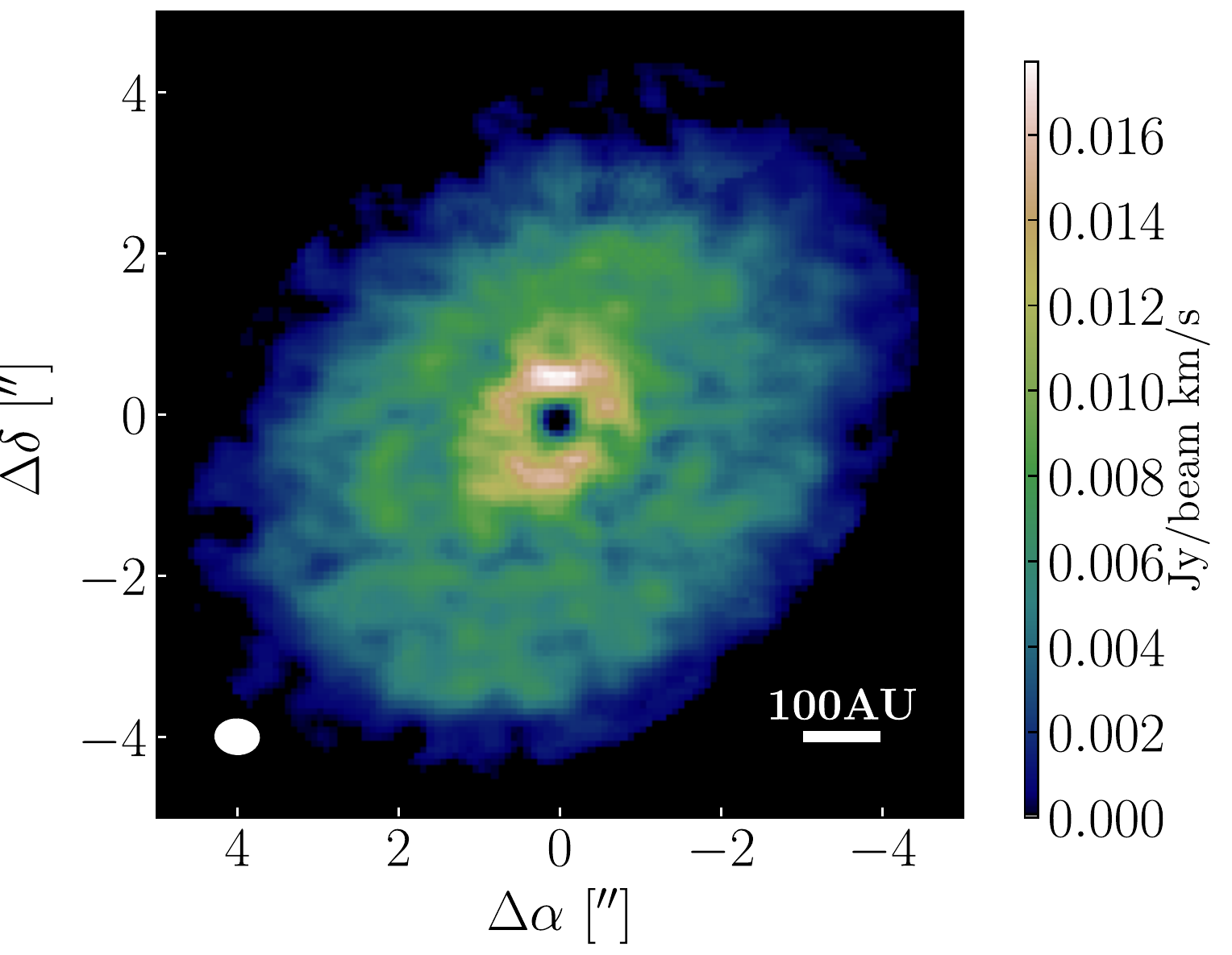}
 \hspace{0.1cm}
 \includegraphics[height=0.25\textwidth]{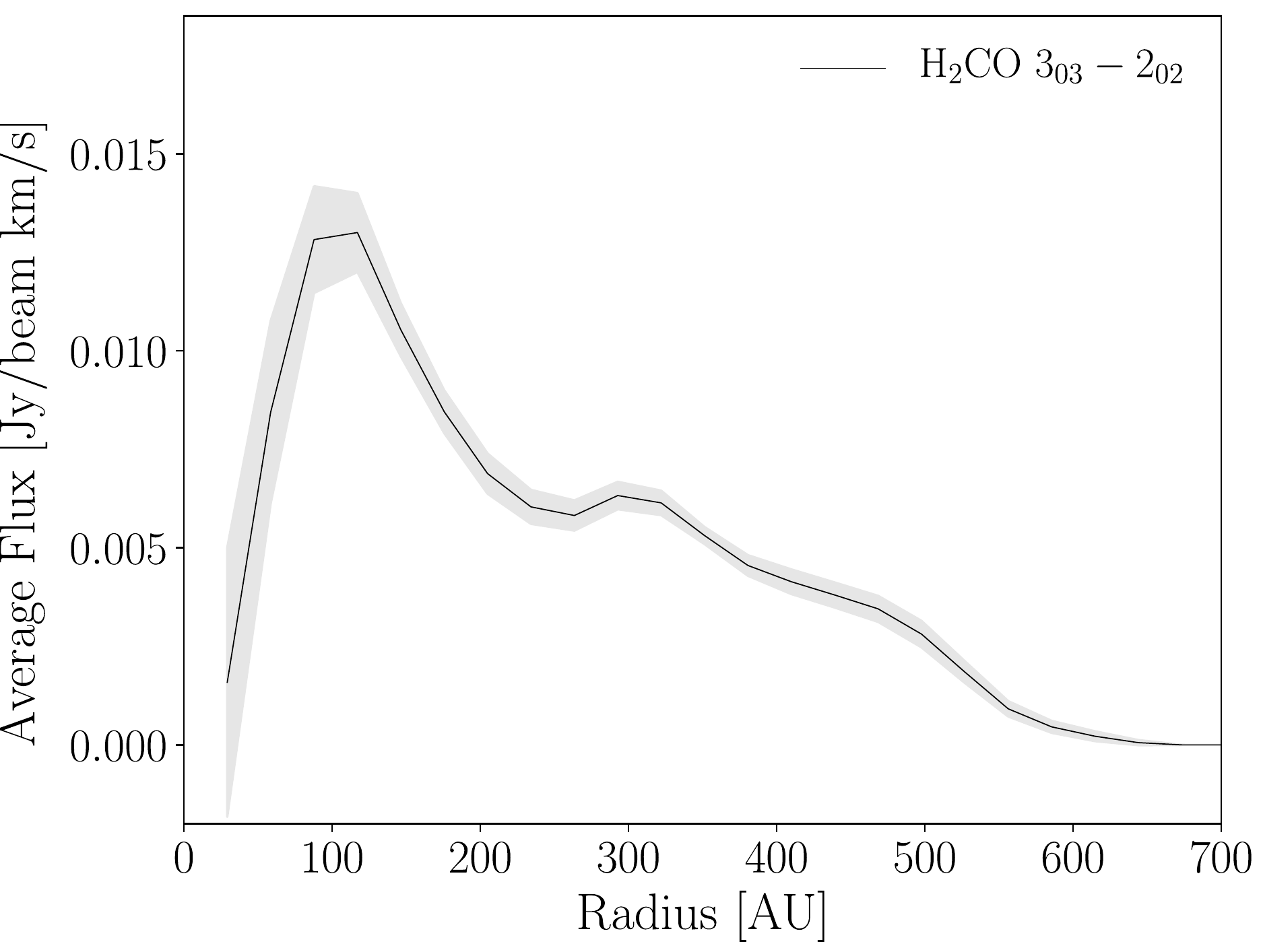}
 \caption{\change{Moment maps and radial profile of H$_2$CO 3$_{03}$--2$_{02}$.
 ({\it Left}) Moment 1 map from 0.76 -- 10.84 km s$^{-1}$,
 clipped at 3$\sigma$. Solid black contours show the 225 GHz/1.3 mm emission at 5.0 $\times$ $10^{-5}$
 (1$\sigma$) $\times$ [5, 10, 25, 50, 100, 300, 500, 1000, 1500, 2000] Jy beam$^{-1}$.
 Synthesized beam and AU scale are shown in the lower corners.
 ({\it Center}) Moment 0 map integrated over 0.76 -- 10.84 km s$^{-1}$ after applying a Keplerian mask. 
 Synthesized beam and AU scale are shown in the lower corners.
 ({\it Right}) Radial intensity curve from azimuthally-averaged elliptical annuli 
 projected to $i$=44$^{\circ}$, P.A.=133$^{\circ}$. Shaded gray area represents 1$\sigma$ errors.}
 }
 \label{fig:mom1}
\end{figure*}

Self-calibration for HD 163296 was performed using the continuum 
TDM spectral window and all line-free channels of the FDM spectral 
windows. DV11 was chosen as the reference antenna. A minimum of 
four baselines per antenna and a minimum signal-to-noise ratio (SNR)
of two were required. Calibration solutions were calculated twice
for phase and once for amplitude. The first phase solution interval 
(solint) was 500s, the second phase and amplitude solutions had solint
equal to the binned integration time (30s). Continuum subtraction of 
the line data was done in the $uv$ plane using 
a single-order polynomial fit to 
the line-free channels. CLEAN imaging was done with natural weighting 
for each continuum-subtracted H$_2$CO line down to a threshold
of 4 mJy. 

This work also uses C$^{18}$O 2--1 \change{calibrated} data of HD 163296 from the 
ALMA \change{project 2011.0.00010.SV obtained from the publicly available
ALMA Science Verification Data website.\footnote{\url{https://almascience.nrao.edu/alma-data/science-verification}}}
See \citet{Rosenfeld2013} for details
on the calibration of the data set. The flux for 
the \change{C$^{18}$O 2--1} line (Table~\ref{tab:obs_par}) 
is consistent with previously reported values \citep{Rosenfeld2013,Qi2015}.
The following software and coding languages were used for 
data analysis in this paper: 
the \textsc{casa} package \citep{McMullin2007}, the \textsc{miriad} 
package \citep{Sault1995}, and \textsc{python}.


\section{Results}
\label{sec:res}

\begin{figure*}[!htbp]
 \centering
 \includegraphics[height=0.5\textheight]{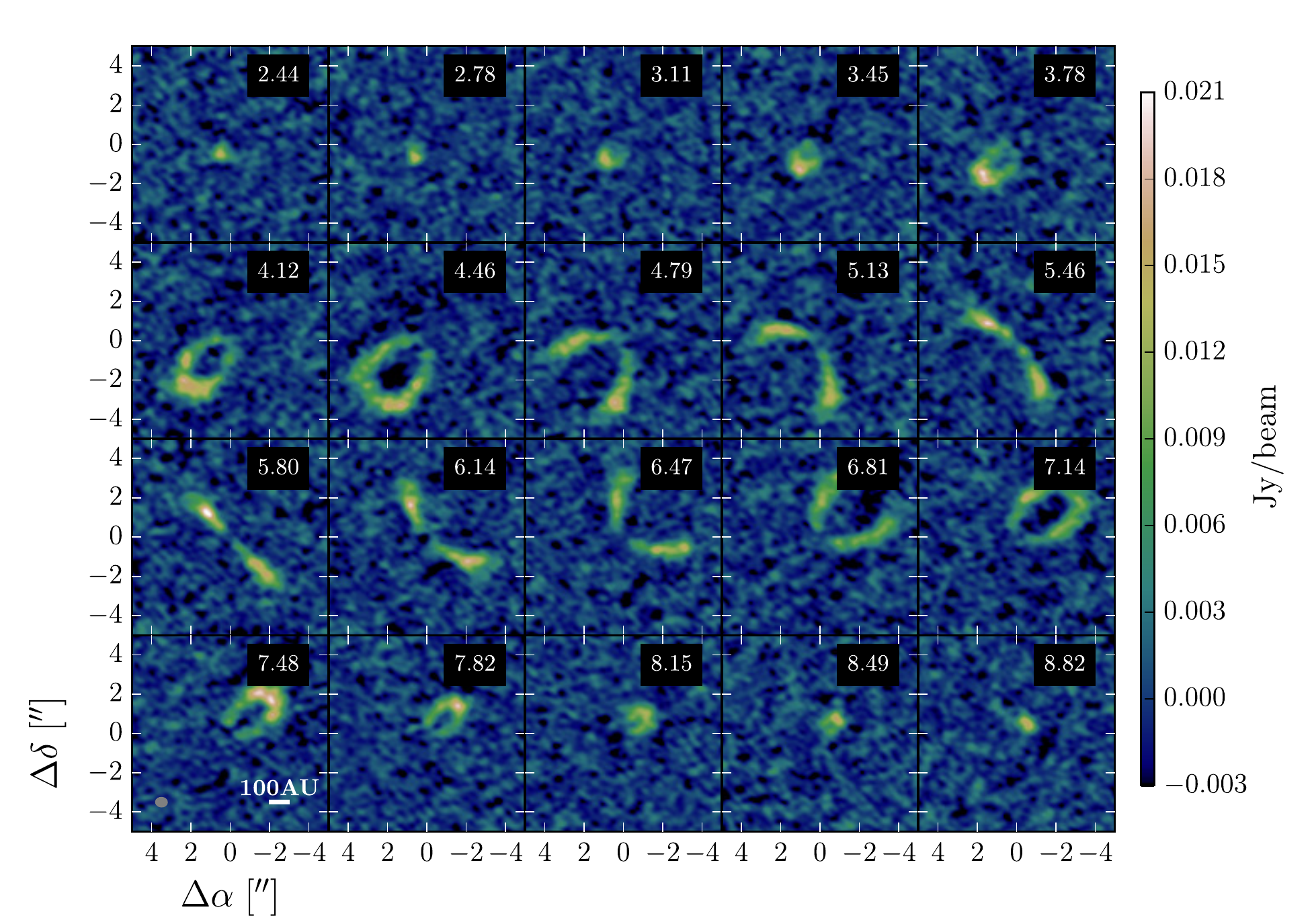}
 \caption{Channel maps of H$_2$CO 3$_{03}$--2$_{02}$ from 2.44--8.82 km s$^{-1}$,
 Hanning smoothed to 0.336 km s$^{-1}$ channels. 
 Channel velocity is shown in the upper-right corner.
 Synthesized beam and AU scale are shown in the lower-left panel.}
 \label{fig:chanmaps}
\end{figure*}

The following sections present results of H$_2$CO observations
in the disk around HD 163296. Physical parameters of the lines 
and their distribution throughout the disk are discussed in 
Section~\ref{sec:res_distr}. Models of H$_2$CO and C$^{18}$O
emission and their abundances are presented in 
Section~\ref{sec:res_mods}.
Constraints on the excitation temperature
of H$_2$CO are discussed in Section~\ref{sec:res_extemp}.

\subsection{Detection and distribution of H$_2$CO}
\label{sec:res_distr}

The spatially integrated spectrum for each H$_2$CO
line can be found in Figure~\ref{fig:spectra}.
\change{The 3$_{03}$--2$_{02}$ transition is
readily detected in the spectrum extracted from
CLEAN imaging. The two weaker lines are not
detected in the extracted spectra, but when applying
a matched-filter technique (see Sec~\ref{sec:res_mf_det}),
the lines are clearly detected and can be used to 
provide constraints on the H$_2$CO excitation temperature}.
Physical parameters of the three lines and the 
continuum can be found in Table~\ref{tab:obs_par}.

HD 163296 has a $V_{\rm LSR}$ systemic velocity of $+$5.8 km s$^{-1}$ \citep{Qi2011}, 
which corresponds well to the central velocity of the 
H$_2$CO 3$_{03}$--2$_{02}$ line.
\change{The H$_2$CO 3$_{03}$--2$_{02}$ line flux was derived after masking
pixels with $<3\sigma$ emission in the image cube. 
The cube was then integrated
spatially over a 7$\arcsec$ radius and over velocity channels 0.76 -- 10.84 km
s$^{-1}$.} Lower limits on H$_2$CO 3$_{22}$--2$_{21}$ and H$_2$CO 3$_{21}$--2$_{20}$
line fluxes are from \change{estimates} via the matched-filter method.
\change{Upper limits on the lines are based on spectra from the CLEAN images
of H$_2$CO 3$_{22}$--2$_{21}$ and H$_2$CO 3$_{21}$--2$_{20}$ and are given at the 3$\sigma_{\rm I}$ 
level, where $\sigma_{\rm I} = 0.5 \sqrt{\pi/{\rm log(2)}} \Delta v \sigma_{\rm rms}$
estimates the area of a Gaussian curve, 
$\Delta v$ is the FWHM of the detected H$_2$CO 3$_{03}$--2$_{02}$, 
and $\sigma_{\rm rms}$ is the rms noise in Jy from the disk-integrated spectra. }

The H$_2$CO 3$_{03}$--2$_{02}$ image has a $0.54\arcsec \times 0.42\arcsec$
[$66 \times 51$ AU] synthesized beam (P.A. = 86.5$^\circ$). 
Figure~\ref{fig:mom1} shows a velocity-weighted 
(first-order moment) map of H$_2$CO 3$_{03}$--2$_{02}$ from 0.76 -- 10.84 km 
s$^{-1}$, clipped at the 3$\sigma$ level, which reveals the full extent of the 
H$_2$CO emission in Keplerian rotation, while Figure~\ref{fig:chanmaps} shows
the channel maps of H$_2$CO 3$_{03}$--2$_{02}$ around HD 163296
\change{Hanning smoothed to a resolution of} 0.336 km s$^{-1}$ over velocities where 
molecular emission is present.
The inner and outer projected radii ($i = 44.0^{\circ}$, P.A. = 133$^{\circ}$
East of North) of H$_2$CO 3$_{03}$--2$_{02}$ emission
at the 3$\sigma$ level along the major axis
are \change{0.4$\arcsec$ and 4.5$\arcsec$}, respectively, corresponding to 
projected physical distances $R_{\rm in} \simeq$ 50 AU and 
$R_{\rm out} \simeq$ 550 AU at a distance of 122 pc \citep{vandenAncker1998}.

The extent of H$_2$CO 3$_{03}$--2$_{02}$ was found to be greater than 
that of the 1.3 mm continuum \change{(shown in 
black contours in Figure~\ref{fig:mom1})}, suggesting that millimeter-sized grains have
decoupled from the gas and drifted radially inward. \citet{deGregorioMonsalvo2013}
observed the same phenomenon in $^{12}$CO and the 850 $\mu$m continuum.
The 1.3 mm continuum has a projected outer radius at 3$\sigma$ of 
2.2$\arcsec$, or $R_{\rm out}^{\rm 1.3mm} \simeq$ 270 AU. The 1.3 mm emission
extends beyond the 850 $\mu$m continuum reported by \citet{deGregorioMonsalvo2013}
due to the increased sensitivity of our observations. \change{\citet{Zhang2016} reported
that analysis of the 1.3 millimeter continuum visibilities in this data set suggests
a ring-like structure not seen in imaging at this resolution. The ring-like
nature of the millimeter dust was confirmed by high-resolution observations
after the original submission of our paper
\citep{Isella2017}. They explained the dust morphology as 
three distinct dust gaps centered at 60, 100, and 160 AU.}

\change{To calculate the H$_2$CO 3$_{03}$--2$_{02}$ radial intensity profile, an 
integrated intensity (zero-order moment) map
was first created by applying a mask in right ascension, declination,
and velocity to the image cube to enhance the signal-to-noise. 
The mask is based on the disk rotational velocity profile, which is
assumed to be Keplerian with a mass of $M = 2.3 \ M_{\odot}$, corresponding
to the mass of the central star. 
In each velocity channel of the image cube, a subset of pixels were chosen
where the calculated Keplerian velocity of the pixels matches the
Doppler-shifted velocity of the line. All pixels that did
not match these criteria were masked. \citet{Yen2016} use a similar
method to extract their integrated intensity maps.}
The radial intensity profile and integrated intensity map for H$_2$CO 3$_{03}$--2$_{02}$ 
emission are shown in Figure~\ref{fig:mom1}.
\change{Azimuthally-averaged elliptical annuli projected to an inclination of 44$^{\circ}$
and position angle of 133$^{\circ}$ were used to calculate the average flux
in each radial step. This method provides more signal-to-noise per annulus, but
results in a decrease in resolution by a factor of two due to the foreshortening 
along the inclined disk's minor axis in our radial intensity profiles. 
Radial step sizes of 0.24$\arcsec$ for H$_2$CO 3$_{03}$--2$_{02}$ 
and 0.4$\arcsec$ for C$^{18}$O 2--1 were used for each annulus to provide a 
sampling of approximately two data points per original beam width.}

\change{The radial profile reveals an absence of emission at the 
center of the disk, a peak in intensity at $\sim$100 AU
with emission then decreasing until a turnover in the profile at $\sim$200 AU
and a bump at $\sim$300 AU, signifying an enhancement in emission in the
outer regions of the disk. 
The same curve for C$^{18}$O has centrally peaked emission and 
intensity decreasing with radius.
Already the shape of the radial profiles
of the two molecules indicates a difference in abundance
gradients throughout the disk. The C$^{18}$O
profile suggests that it follows more or less the smoothly
decreasing H$_2$ gas 
density. On the other hand, H$_2$CO shows a peak at the
approximate location of the CO snow line at 90 AU \citep{Qi2015}, and 
another enhancement is located roughly at the edge of the dust continuum.
Such a radial profile highlights the need for two H$_2$CO formation
mechanisms to account for the observed emission: one warm route
that produces emission at temperatures above that of CO freeze-out
in the inner disk within 100 AU
and one cold route that produces emission outside of the CO snow line.
Further explanations for these features are given in 
Section~\ref{sec:res_mods} and Section~\ref{sec:disc}.}

\subsubsection{Matched filter detections}
\label{sec:res_mf_det}

After subtracting the continuum from the line data, we employed a matched filter 
technique to the visibilities to detect the
weaker H$_2$CO 3$_{22}$--2$_{21}$ and H$_2$CO 3$_{21}$--2$_{20}$ lines. 
\change{In this technique, an image cube containing a template emission profile is 
sampled in $uv$ space to obtain a set of template visibilities that act as
the filter. The template
is then cross-correlated with a set of visibilities with a low signal-to-noise ratio 
(SNR) in an attempt to detect the presence of the template emission within the low SNR data set.
The cross-correlation is done by sliding the template visibilities channel-by-channel
across the velocity axis of the low SNR visibilities. When the template reaches
the source velocity in the low SNR data, there will be a sharp peak in the filter 
response spectrum of 
the correlation if the template signal is detected within the low SNR visibilities. 
This is analogous to image-based stacking approaches \citep[e.g.][]{Yen2016}, 
but retains the advantages of working in the $uv$ plane.}
In this work, to obtain a data-based \change{template} for the matched filter method, the 
H$_2$CO 3$_{03}$--2$_{02}$ line was re-imaged with CLEAN in 
0.084 km s$^{-1}$ velocity channels using a $uv$ taper to achieve a 
1$\arcsec$ synthesized beam. Image channels showing H$_2$CO emission (1.6--10 km s$^{-1}$) 
were sampled in the $uv$ plane using the \textsc{python}
\texttt{vis\_sample}\footnote{\texttt{vis\_sample} is publicly 
available at \url{https://github.com/AstroChem/vis\_sample} or in the 
Anaconda Cloud at \url{https://anaconda.org/rloomis/vis\_sample}} routine, and the resulting 
visibilities were then used as the \change{template} signal. 

\begin{figure}[!b]
 \centering
 \includegraphics[width=0.48\textwidth]{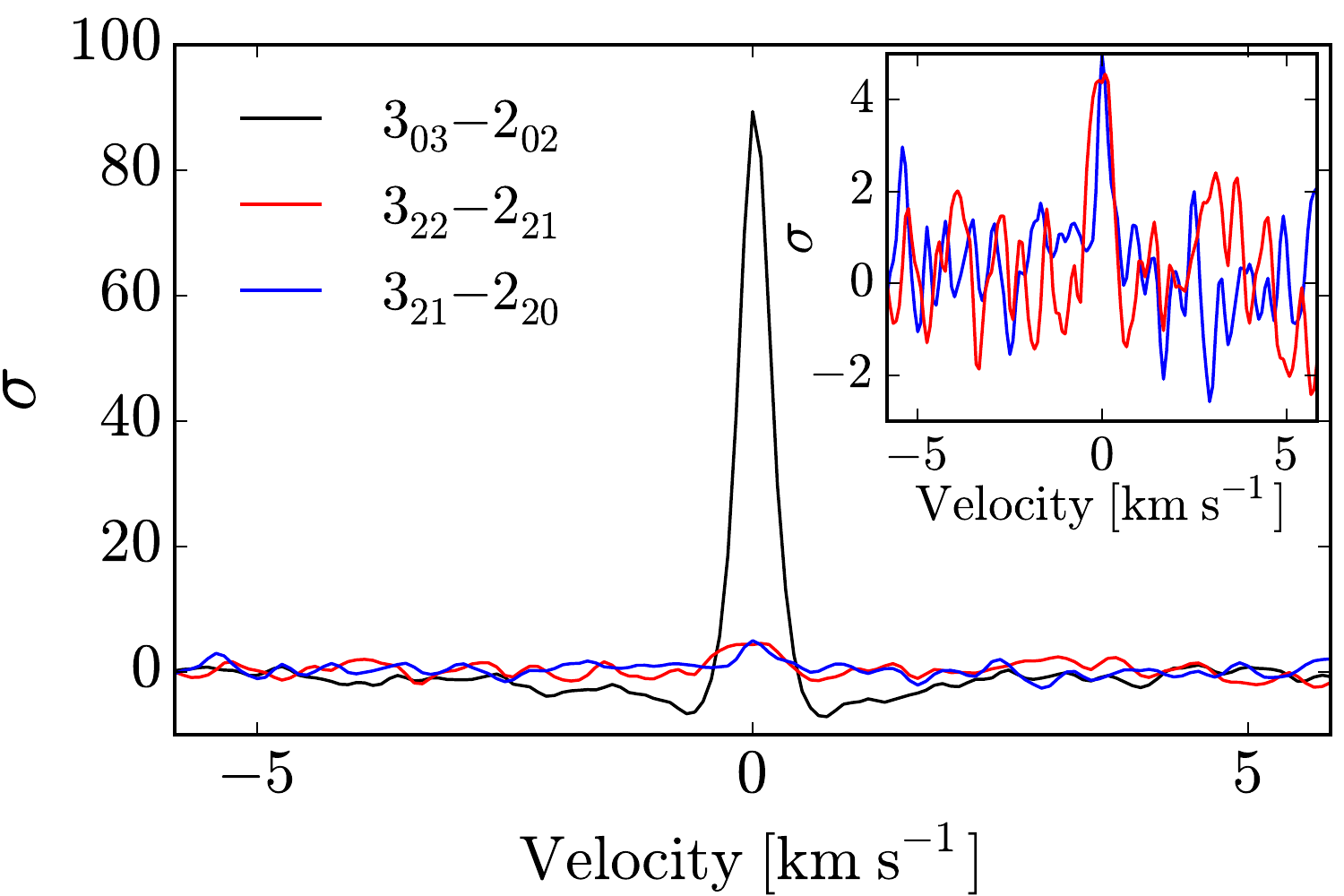}
 \caption{Matched filter responses of the observed H$_2$CO lines to the 
 H$_2$CO 3$_{03}$--2$_{02}$ data-based \change{template}. Self-response (black) shows
 \change{template} recovery of the 3$_{03}$--2$_{02}$ detection. \textit{Inset:} H$_2$CO 3$_{22}$--2$_{21}$ (red)
 and H$_2$CO 3$_{21}$--2$_{20}$ (blue) are detected at the 4.5$\sigma$ and 5$\sigma$ level, respectively.}
 \label{fig:h2cofilter}
\end{figure}

\change{Figure~\ref{fig:h2cofilter} shows the filter impulse responses of the three H$_2$CO 
visibility data sets to the H$_2$CO 3$_{03}$--2$_{02}$ \change{template}. The black curve is
the response of the H$_2$CO 3$_{03}$--2$_{02}$ visibility data to the \change{template}, highlighting
the effectiveness of the filter to recover the line detection. The inset reveals the 
4.5$\sigma$ and 5$\sigma$ detections of H$_2$CO 3$_{22}$--2$_{21}$ and H$_2$CO 
3$_{21}$--2$_{20}$, respectively, 
where $\sigma$ is calculated as the standard deviation of the response of emission-free 
visibility channels to the \change{template}. To constrain the total flux of the weaker lines, 
we compare the ratio of their peak filter responses and the peak response of 
the H$_2$CO 3$_{03}$--2$_{02}$ visibilities (90$\sigma$, Figure~\ref{fig:h2cofilter}).
Under the assumption that all three observed H$_2$CO lines are co-spatial, 
the $\sigma$-ratio can be used to estimate the weaker line fluxes 
reported in Table ~\ref{tab:obs_par}. 
The response of the \change{template} is limited by how well it spatially matches the emission, 
thus making the derived line fluxes lower limits.}

\subsection{Modeling H$_2$CO and C$^{18}$O emission}
\label{sec:res_mods}

Previous studies \citep{Qi2011,Rosenfeld2013,Qi2015} have attempted to use CO
isotopologues to determine the radial location of CO freeze-out in HD 163296.
\citet{Qi2011} modeled the $^{13}$CO isotope and found a distinct drop in
abundance at $\sim$155 AU, which they attributed to CO freeze-out.
However, in \citet{Qi2015} they claim $^{13}$CO is a less robust
tracer as it is difficult to separate CO freeze-out from opacity effects.
$^{13}$CO may remain optically thick out to radii beyond 100 AU. Thus, 
the apparent depletion may be due to a decrease in optical depth rather than
an actual drop in abundance. They use 
C$^{18}$O as a more robust, optically thin tracer of the column density
of CO throughout the disk. Following this reasoning, we model only the 
C$^{18}$O isotopologue to reveal structure in the CO gas. 
\change{Although the C$^{18}$O Science Verification data has been 
previously reported \citep{Rosenfeld2013,Qi2015}, we reanalyze the data in an effort
to provide a ground truth for the CO surface density --
particularly for the outer disk -- within the same modeling approach 
as used for H$_2$CO and within the limits of 
the data resolution and our disk model.} 

The aims of modeling H$_2$CO and C$^{18}$O are to determine likely
formation scenarios for H$_2$CO and any relation to the CO snow line.
\change{If H$_2$CO is abundant in regions close to or below the CO freeze-out
temperature, grain surface formation of H$_2$CO on CO ices will contribute.
If H$_2$CO is abundant only at high temperatures zones of the disk, then
gas-phase production of H$_2$CO dominates.} 
By varying the relative molecular abundances in different regions
of the models and comparing the model distribution
to the data, we can determine which parts of the disk are harboring 
reservoirs of H$_2$CO. 

This section describes the models used to reproduce 
the observed H$_2$CO 3$_{03}$--2$_{02}$ and C$^{18}$O 2--1 emission
based on the HD 163296 disk model created by \citet{Qi2011}.
In their paper they constrain the radial and vertical density and temperature 
structure of a steady viscous accretion disk with an exponentially-tapered 
edge. Fitting the model continuum at multiple wavelengths to the observed SED constrained
the radial structure. Observations of multiple optically thick $^{12}$CO transitions 
were used to constrain the vertical structure. \change{A modified version of this} physical
model was used by \citet{Mathews2013} to determine the distribution
of DCO$^{+}$ in HD 163296. \change{To constrain the vertical structure of the 
dust in our physical model, \citet{Mathews2013} 
refit the SED by varying independently the dust scale heights 
of Gaussian distributions of small ($a_{\rm max}$ = 25 $\mu$m) 
and large ($a_{\rm max}$ = 1 mm) populations of dust grains. 
Similarly, the vertical gas density distribution is treated as a 
two-component model with independent scale heights
to simulate a Gaussian distribution
at low heights with an extended tail higher in the disk.
The gas scale heights are varied to recover the CO fluxes
reported in \citet{Qi2011}.
Given these dust and gas distributions and
assuming the dust continuum to be optically thin,
the gas surface density of both H$_2$CO and C$^{18}$O
should be robustly measured in our models.} 

\change{In this work,} the \citet{Mathews2013} model 
was used as the physical disk structure for simulating molecular emission
using the LIne Modeling Engine \citep[LIME,][]{Brinch2010}
3D radiative transfer code. Synthesized data cubes were created with LIME for 
H$_2$CO 3$_{03}$--2$_{02}$ and C$^{18}$O 2--1 in non-LTE with H$_2$ as the primary collision partner.
Both ortho- and para-H$_2$ species were included in collisional excitation,
with a temperature-dependent ortho- to para- ratio (OPR) such that OPR = 3 at
temperatures $\geq$200 K and decreases exponentially at lower temperatures.
Molecular collision rates were taken from the Leiden Atomic and Molecular
Database \citep[LAMDA,][]{Schoeier2005}. 
The disk inclination, position angle, and distance are set to $i$ = 44.0$^{\circ}$, 
P.A. = 133.0$^{\circ}$, and $d$ = 122 pc. 

\begin{figure}[!t]
 \centering
 \includegraphics[width=0.5\textwidth]{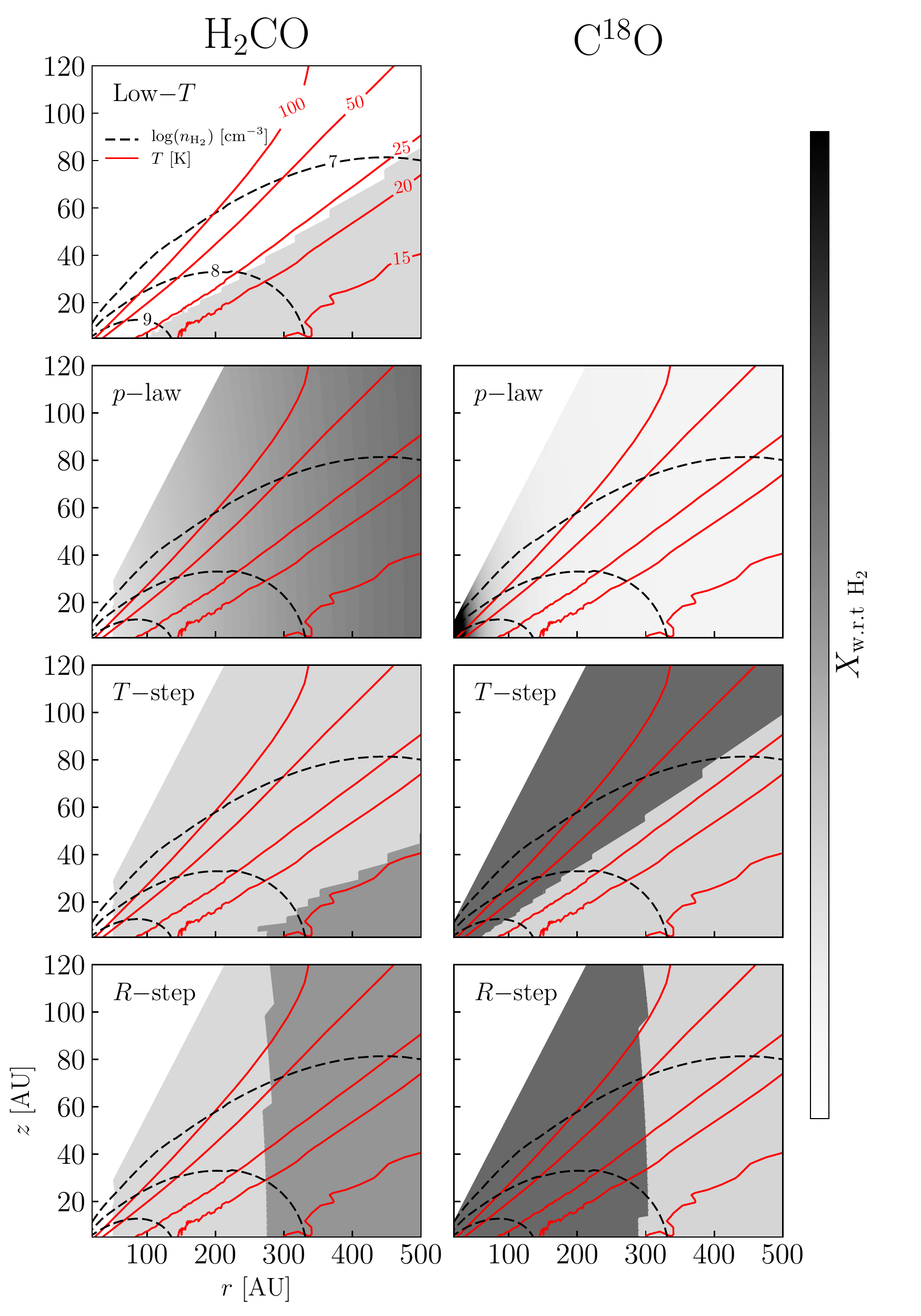}
 \caption{\change{Toy model abundance scenarios for H$_2$CO (left) and C$^{18}$O (right). 
 $X$ is the molecular abundance with respect to molecular hydrogen (grayscale). 
 Red solid contours show the temperature structure of the gas in the disk. 
 Black dashed contours show the density structure of the gas in the disk as
 the log of the molecular hydrogen number density. The $X$
 distribution in each panel follows the 
 best-fit normalized model (Table~\ref{tab:mod_par_norm}).}
 }
 \label{fig:toymodel}
\end{figure}

Four types of models are used to test
the distribution of observed H$_2$CO 3$_{03}$--2$_{02}$
with different fractional abundance profiles relative to H$_2$. 
Figure~\ref{fig:toymodel} depicts examples of each of these scenarios
with the relevant disk regions.
Three of these models are used for C$^{18}$O 2--1. The first model assumes
a constant abundance constrained to low temperatures where H$_2$CO
formation on the surface of icy grains is favorable (Section~\ref{sec:res_mods_cofreeze}).
The low-temperature model is not used for C$^{18}$O 2--1. 
In the second model, H$_2$CO 3$_{03}$--2$_{02}$ and C$^{18}$O 2--1 
have a power-law abundance profile (Section~\ref{sec:res_mods_plaw}). 
The third model has a temperature-based step-abundance profile 
with a constant inner (high-temp) and outer (low-temp) abundance 
and a change-over temperature $T_{\rm c}$ as the boundary (Section~\ref{sec:res_mods_2phase}).
The final model has a radial step-abundance profile with a constant 
inner abundance, \change{constant} outer abundance, and change-over radius $R_{\rm c}$ (Section~\ref{sec:res_mods_step}).
Analysis of the models makes use of \change{the \texttt{vis\_sample} routine
to read the $uv$ coordinates directly from an observed ALMA measurement set
and create synthetic visibilities based on an input sky model.}

\change{A central hole is observed in the H$_2$CO data, as 
seen in Figure~\ref{fig:mom1}, with a size approximately
equal to the width of the beam.}
This hole is likely a result of strong absorption by an 
optically thick dust continuum (see also Section~\ref{sec:disc_h2co_hole}). 
\change{Beyond 50 AU, the optical depth radial profile for the 
LIME model continuum is found to be optically thin with 
$\tau < 0.6$, which ensures that features in the gas radial
profile outside of 50 AU are not caused by dust opacity effects. 
The inner region
($<$50 AU) cannot be properly modeled here due to the low resolution
of the observations, which do not allow for 
proper description of any dust substructure. The modeling of
\citet{Zhang2016} and new high-resolution observations by 
\citet{Isella2017} show significant substructure
in the dust and a large increase in optical depth in
the inner 50 AU. Such substructure is unlikely to be accurately
described in our models, thus we ignore radii $<$50 AU.} The central hole is 
therefore treated as an H$_2$CO abundance inner radius in the modeling.
  
All H$_2$CO 3$_{03}$--2$_{02}$ models have an inner radius set to $R_{\rm in}$ = 50 AU.
$R_{\rm in}$ was constrained for H$_2$CO by varying the inner radius of a
constant abundance model to determine the best fit to the 
inner 150 AU of the radial intensity curve.
Thereafter, $R_{\rm in}$ remains a fixed parameter in the models.
C$^{18}$O 2--1 models have no such inner radius as 
the emission is centrally peaked.

Each LIME model was continuum-subtracted before running \texttt{vis\_sample}.
\change{We first tested H$_2$CO 3$_{03}$--2$_{02}$ and C$^{18}$O 2--1
models normalized to the total flux of the data in order to find
the best-fit to the spatial distribution of each line, then we varied the
abundance of the best-fit normalized model to match the absolute flux of the data. 
To determine the total flux, we took a vector average of visibilities with 
baselines $<$30 m and integrated over all channels containing emission. The
model was then scaled to match the total flux of the data.
Goodness of fit for each model was determined by $\chi^2$ minimization
between the normalized visibilities of the model and the visibilities of the data.
Initial reference abundances were chosen for the normalized models
to ensure optically thin line emission.}
The H$_2$CO reference abundance was set to $X = 1.0 \ \times \ 10^{-12}$. 
C$^{18}$O models used a reference abundance 
of $X = 1.0  \ \times \ 10^{-7}$. All normalized models
remained optically thin \change{with $\tau < 1$}; there was no significant 
increase in the optical depth profile \change{of the models for the
parameter space explored here.
The best-fit normalized model for each line was then used to 
vary the molecular abundances to find the best agreement 
between the absolute flux of the model and of the data using $\chi^2$ minimization
on the visibilities.}

\begin{figure*}[!htbp]
 \raggedleft 
 \includegraphics[width=0.45\textwidth]{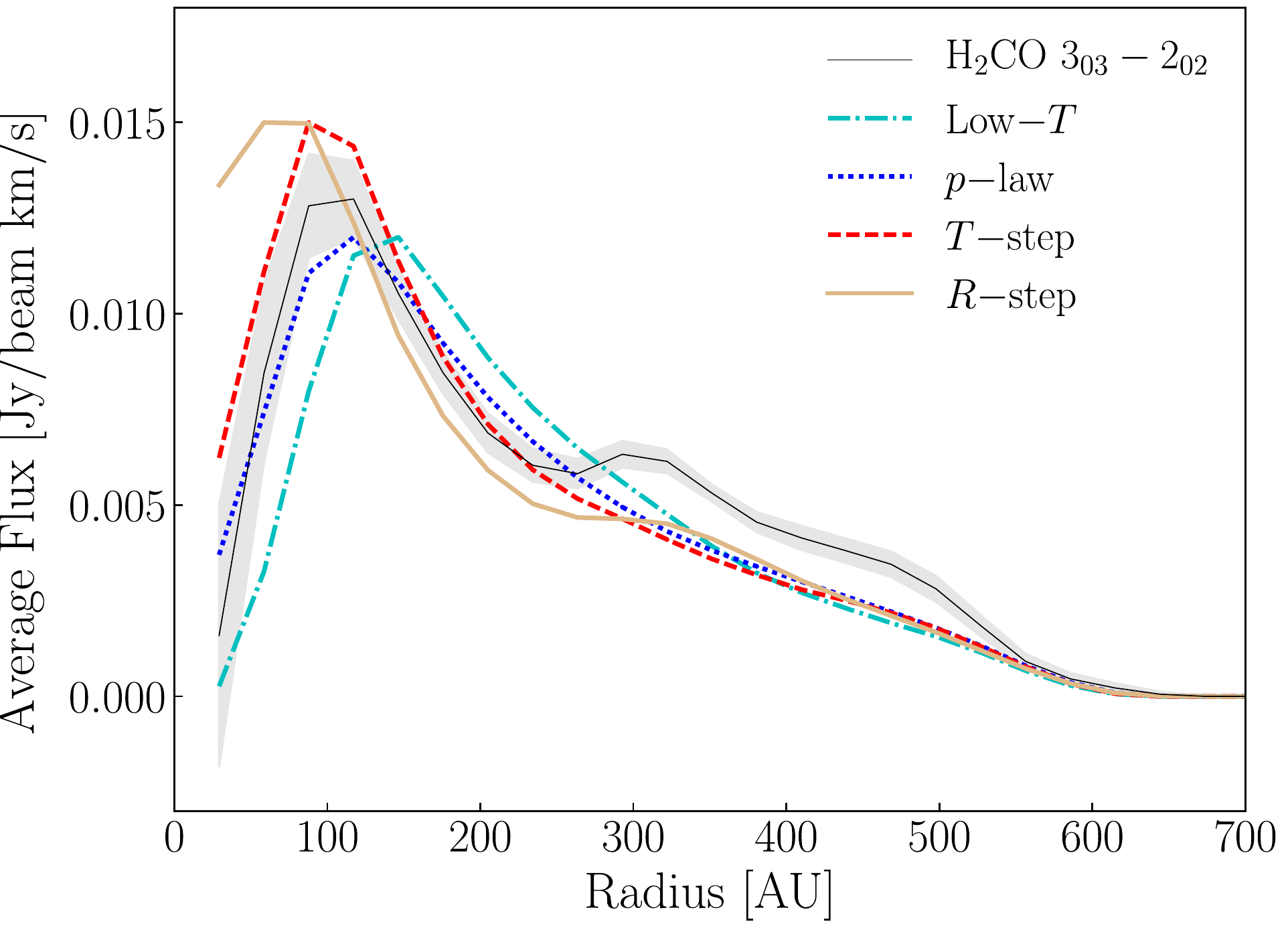}
 \hspace{0.5cm}
 \centering
 \includegraphics[width=0.45\textwidth]{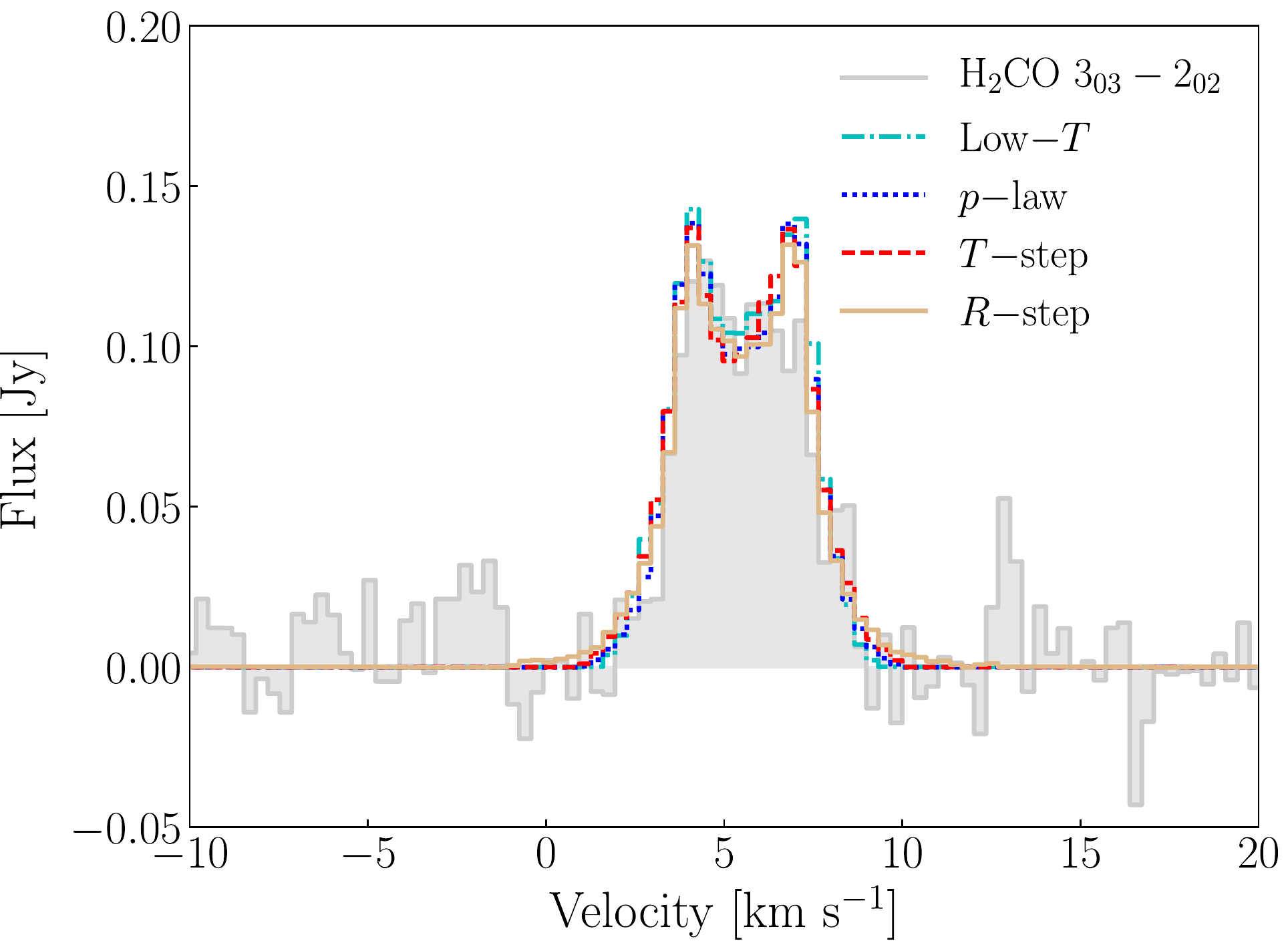} \\ 
 \raggedleft
 \hspace{-0.2cm}
 \includegraphics[width=0.435\textwidth]{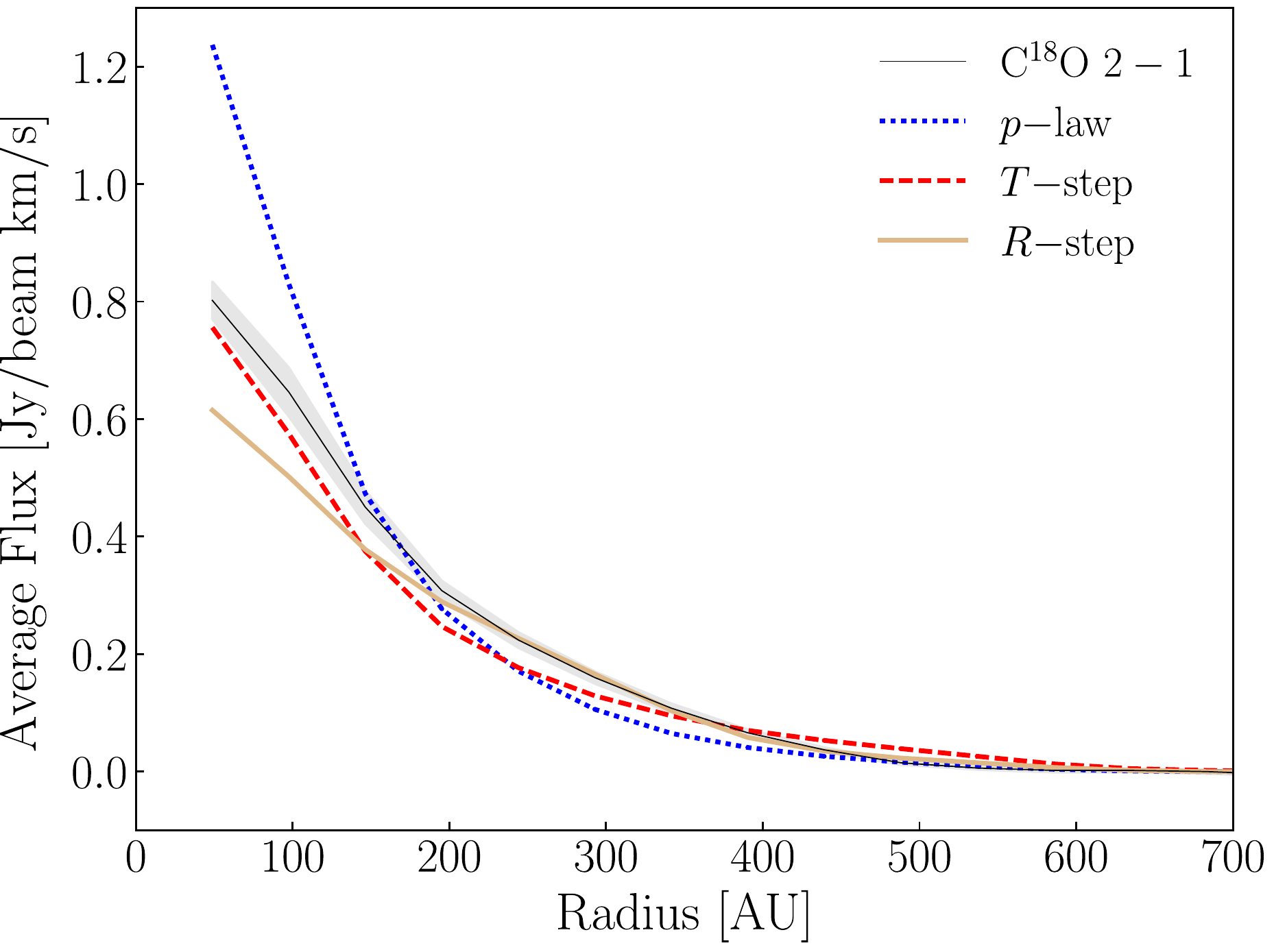}
 \hspace{0.9cm}
 \centering
 \includegraphics[width=0.43\textwidth]{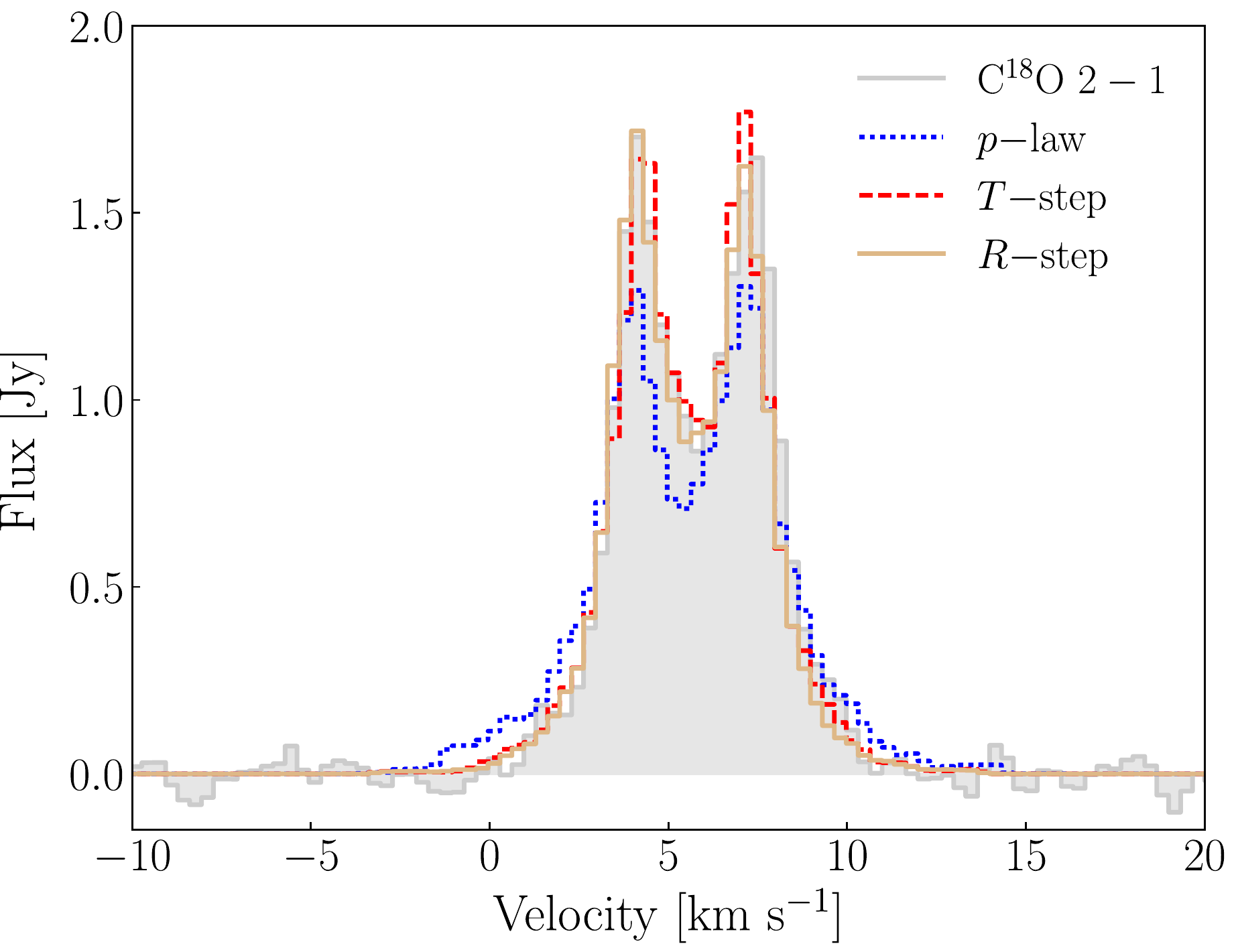}
 \hspace{-0.6cm}
 \caption{ H$_2$CO 3$_{03}$--2$_{02}$ 
 and C$^{18}$O 2--1 data are compared with best-fit normalized
 models for each scenario mentioned in Section~\ref{sec:res_mods}. The low-temperature model 
 (dot-dashed cyan), the power-law model (dotted blue), the temperature step-abundance 
 model (dashed red), and the radial step abundance model (solid gold) show 
 the radial distribution \change{and spectra}. 
 {\it (Left)} Radial intensity curves \change{of the best-fit normalized models}
 obtained from \change{azimuthally-averaged elliptical annuli projected to $i$=44$^{\circ}$, P.A.=133$^{\circ}$. 
 The shaded grey region represents 1$\sigma$ error bars.
 H$_2$CO profiles are taken from integrated intensity maps after applying a Keplerian mask.}
 {\it (Right)} \change{Disk-integrated spectra of the best-fit normalized models obtained from 
 a 5.6$\arcsec$ circular aperture. H$_2$CO spectra are Hanning smoothed to 0.336 km $^{-1}$ channels.}
 Parameters for
 each model can be found in Table~\ref{tab:mod_par_norm}.
 }
 \label{fig:normmods}
\end{figure*}

\begin{table}[!htbp]
 \caption{Best-Fit Normalized Models}
 \centering
 \label{tab:mod_par_norm}
 \resizebox{9cm}{!}{
 \begin{tabular}{lccccc}
 \hline \hline
 \multicolumn{6}{c}{H$_2$CO 3$_{03}$--2$_{02}$} \\
 \hline 
 Abundance Model & $p$ & $T_{\rm c}$ [K] & $R_{\rm in}$\tablefootmark{a} [AU] & $R_{\rm c}$ [AU] & $X_1/X_2$  \\
 \hline
 Low-temperature & -- & 24$\pm$2 & -- & 65$\pm$15\tablefootmark{$\dagger$} & --  \\
 Power-law & $+$0.5 & -- & 50 & -- & --  \\
 Temperature step & -- & 16$\pm$2 & 50 & 230$\pm$60\tablefootmark{$\dagger$} & 0.5  \\
 Radial step & -- & 15$\pm$1\tablefootmark{$\dagger$} & 50 & 270$\pm$20 & 0.5  \\
 \hline \hline
 \multicolumn{6}{c}{C$^{18}$O 2--1} \\
 \hline
 Abundance Model & $p$ & $T_{\rm c}$ [K] & $R_{\rm in}$\tablefootmark{a} [AU] & $R_{\rm c}$ [AU] & $X_1/X_2$  \\
 \hline
 Power-law & $-$2 & -- & 0.1 & -- & --  \\
 Temperature step & -- & 32$\pm$2 & 0.1 & 32$\pm$5\tablefootmark{$\dagger$} & 10  \\
 Radial step & -- & 15$\pm$1\tablefootmark{$\dagger$} & 0.1 & 290$\pm$20 & 10  \\
 \hline
 \end{tabular}
 }
 \tablefoot{$\chi^2$ values are reduced by the number of points and free parameters in each model.
 \tablefoottext{a}{Fixed parameter.}
 \tablefoottext{$\dagger$}{Indicates the corresponding midplane value to the best-fit model
 parameter based on the density and temperature structure of the \citet{Mathews2013} physical model.}
 }
\end{table}

\subsubsection{Low-temperature abundance model}
\label{sec:res_mods_cofreeze}

The low-temperature model simulated H$_2$CO emission
that is present due to grain-surface chemistry 
in regions below the expected CO 
freeze-out temperature \change{and subsequent non-thermal
desorption from icy grains.} The models used a constant fractional abundance 
relative to H$_2$, constrained by a threshold
temperature. Above the threshold temperature the H$_2$CO
abundance was set to zero everywhere. Based on estimates of CO
freeze-out temperatures from \citet{Qi2015},
model threshold temperatures range from
14--50 K in steps of 2 K. 
Below the threshold temperature, gas-phase H$_2$CO \change{is present}. 
It is assumed that there is a mechanism to
stimulate sufficient desorption of H$_2$CO from the 
icy grains, such as UV or X-ray photodesorption, or cosmic rays 
penetrating the disk midplane.

The best fit for the normalized low-temperature model
for H$_2$CO has a threshold temperature of 24$\pm$2 K, corresponding
to a midplane radius of 65$\pm$15 AU. Seen in Figure~\ref{fig:normmods}, 
the model radial intensity curve \change{fails to recover the 
sharp decrease in emission between 100--200 AU and 
the turnover and secondary bump beyond $\sim$200 AU.}
It is clear that a scenario in which H$_2$CO originates entirely 
beyond the CO freeze-out temperature is not a good representation of
the distribution seen in the observations. There must be H$_2$CO present in
other parts of the disk.

\subsubsection{Power-law abundance model}
\label{sec:res_mods_plaw}

In these models a varying abundance profile was considered
for both H$_2$CO and C$^{18}$O,
following a power-law distribution,

\begin{equation*}
 X = X_{\rm 100AU} \left(\frac{R}{\rm 100 \ AU}\right)^{p},
\end{equation*}

\noindent where $X_{\rm 100AU}$ is the abundance at 100 AU,
$R$ is the disk radius, and $p$ is the
power-law index. C$^{18}$O is present throughout the disk. H$_2$CO has
an inner radius $R_{\rm in}$ = 50 AU, which was used in
all subsequent H$_2$CO modeling.

The best-fit power-law H$_2$CO model has $p = 0.5$, \change{with the abundance increasing
with radii}. The best-fit value found here is \change{more gradual than} the $p = 2$ positive
power law slope found by \citet{Qi2013}, \change{but both suggest that there is} increased
H$_2$CO production occurring in the outer disk. However, the $p = 0.5$ model does not 
provide the overall normalized best fit to the H$_2$CO 3$_{03}$--2$_{02}$ data presented here,
as seen in Figure~\ref{fig:normmods}. The best-fit C$^{18}$O  model had $p = -2$, suggesting C$^{18}$O is 
centrally peaked, but with a decreasing abundance in the outer regions of the disk.
The model radial intensity curve under-produces emission \change{beyond 200 AU}
and overproduces emission \change{inside of 200 AU}.

The simple power-law model does not capture the distribution seen in
either H$_2$CO or C$^{18}$O. The failure 
of the H$_2$CO model to recover the
\change{shape of the radial intensity profile suggests} that there are changes
in the distribution of emission not
captured in this model; we are underestimating the contribution from grain
surface formation. The failure of the C$^{18}$O power-law model indicates 
that the effect of CO \change{depletion} is not properly taken into account.
\change{To reproduce the data at our resolution, the C$^{18}$O abundance profile
needs an abrupt change rather than the gradual change provided by the power-law model.}

\subsubsection{Temperature step-abundance model}
\label{sec:res_mods_2phase}

Two-phase abundance models with a change-over 
temperature that distinguishes between the warm and cold 
regions of the disk were created to test H$_2$CO formed in the
gas phase and H$_2$CO originating from icy grains, respectively. 
We assume that the change-over temperature represents
\change{the boundary} below which H$_2$CO should form 
via hydrogenation of CO ice. \change{The temperature step-abundance
model for C$^{18}$O reflects the freeze out of CO, 
both radially and vertically, since there also is a vertical 
temperature gradient. While in these models we parameterize 
the C$^{18}$O abundance with a change-over temperature, it is important 
to remember that this results in a radial column density 
profile that decreases gradually and extends well beyond the 
midplane CO snow line. Given our limited angular resolution, 
our data primarily samples the radial extent of the disk surface 
layer where C$^{18}$O is present in the gas phase. Although we parameterize 
this with a temperature, we caution against the simplistic 
interpretation as an evaporation temperature, since its value depends 
on how well we know the vertical temperature structure and because 
our data do not resolve the location of the midplane CO snow line.}

The change-over temperature $T_{\rm c}$ was tested in the range 
12 -- 36 K in steps of 2 K. The abundance ratio between
the inner and outer 
regions varies to cover the range
$X_1/X_2$ from 0.001 -- 10 (0.1 -- 1000 for C$^{18}$O).
The best-fit
H$_2$CO model has a change-over temperature $T_{\rm c} = 16 \pm 2$ K and an abundance
ratio $X_1/X_2 = 0.5$, indicating a \change{factor of 2} enhancement of 
H$_2$CO in the outer regions, but the model does not reproduce
emission beyond 200 AU well (Figure~\ref{fig:normmods}). 
C$^{18}$O is best fit by a change-over temperature $T_{\rm c} = 32 \pm 2$ K with an order of magnitude
reduction ($X_1/X_2 = 10$) in the outer regions.
The temperature step-abundance model provides an improved normalized fit
to the C$^{18}$O observational data over models 1 \& 2 and is consistent with
CO \change{depletion in the cold, outer disk.} 

\change{As explained above, we do not claim that T$_{\rm c}$ is the evaporation 
temperature of CO, but rather that the value of T$_{\rm c}$ results in a 
reasonable match of the radial column density distribution of C$^{18}$O 
given our adopted temperature structure and the limited angular 
resolution of the data. Even then,} the radial profile
of this model under-produces C$^{18}$O within $\sim$400 AU and 
overproduces C$^{18}$O outside of $\sim$400 AU. 

While this model provides a better fit to the H$_2$CO emission than models 1 \& 2, 
it fails to recover the shape of the turnover in the radial profile seen at $\sim$200 AU. 
\change{Instead, the temperature-based boundary causes a gradual change in the radial
intensity due to the vertical temperature structure in the disk. To better fit
the turnover seen in the radial profile, the H$_2$CO abundance profile must have an even more
abrupt radial change. The improvement of the C$^{18}$O normalized model fit over models 1 \& 2
suggests CO freeze-out in the cold, outer parts of the disk.}

\subsubsection{Radial step-abundance model}
\label{sec:res_mods_step}

In these models, 
molecular gas abundance is constant throughout the vertical extent 
of the disk with different abundance values in the inner and outer
regions across the change-over radius.
The outer abundance was varied such that $X_1/X_2$ 
spanned 0.1 -- 10 for H$_2$CO and 0.1 -- 1000 for C$^{18}$O.
The change-over radius $R_{\rm c}$ ranged from 
210 -- 410 AU for H$_2$CO and 70 -- 350 for C$^{18}$O
in steps of 20 AU.

The radial step-abundance model \change{reproduces the turnover 
seen in the radial intensity of the H$_2$CO emission better than the first
three models}. Best-fit parameters are
a change-over radius \change{$R_{\rm c}$ = 270$\pm$20 AU and an abundance ratio
$X_1/X_2 = 0.5$. The radial step-abundance model gives a radial intensity
profile that has a steep drop between $\sim$100--200 AU and a sharp turnover
and plateau beyond $\sim$200 AU.}
Best-fit C$^{18}$O models have $R_{\rm c}$ = 290$\pm$20 AU and 
$X_1/X_2 = $10, indicating a factor of ten depletion of CO in the outer disk
beyond the edge of the millimeter grains. 
\change{This abundance scenario also provides a better normalized fit
than models 1 \& 2, and reproduces the distribution of C$^{18}$O as well
as model 3.}

\change{In this model} the H$_2$CO bump in the radial intensity curve 
is well-captured due to the sharp change in abundance across
the change-over radius. The radial step-abundance model provides the
right amount of H$_2$CO production in the inner and outer regions, \change{likely a
combination of gas-phase and grain-surface formation. 
Penetrating UV photons could photodesorb H$_2$CO that has formed via hydrogenation
of CO ices beyond $\sim$300 AU.
There may also be more gas-phase H$_2$CO formation beyond the edge of the 
millimeter continuum at $\sim$270$\pm$20 AU if UV photons can photodissociate
CO in the upper layers and activate hydrocarbon chemistry for a more efficient CH$_3$ + O pathway.
The C$^{18}$O radial step-abundance model provides an alternative scenario 
for outer disk CO depletion compared to model 3. If the micron-sized
grains are depleted in the outer disk similar to the millimeter-sized grains,
UV photons could photodissociate CO beyond $\sim$300 AU.} 

\change{The fact that both a radial step at 290$\pm$20 AU and a temperature step at 32$\pm$2 K
(32$\pm$5 AU near the midplane) equally well fit the C$^{18}$O data underlines our caution
against interpreting the value of $T_{\rm c}$ as the evaporation temperature. 
The consequences of each model scenario are further discussed in Section~\ref{sec:disc}.
The radial step-abundance case is chosen 
as the C$^{18}$O normalized model for estimating abundances in Section~\ref{sec:res_abun}.}

\subsubsection{H$_2$CO and C$^{18}$O abundance}
\label{sec:res_abun}

\begin{figure*}[!t]
 \raggedleft
 \includegraphics[width=0.45\textwidth,height=6cm]{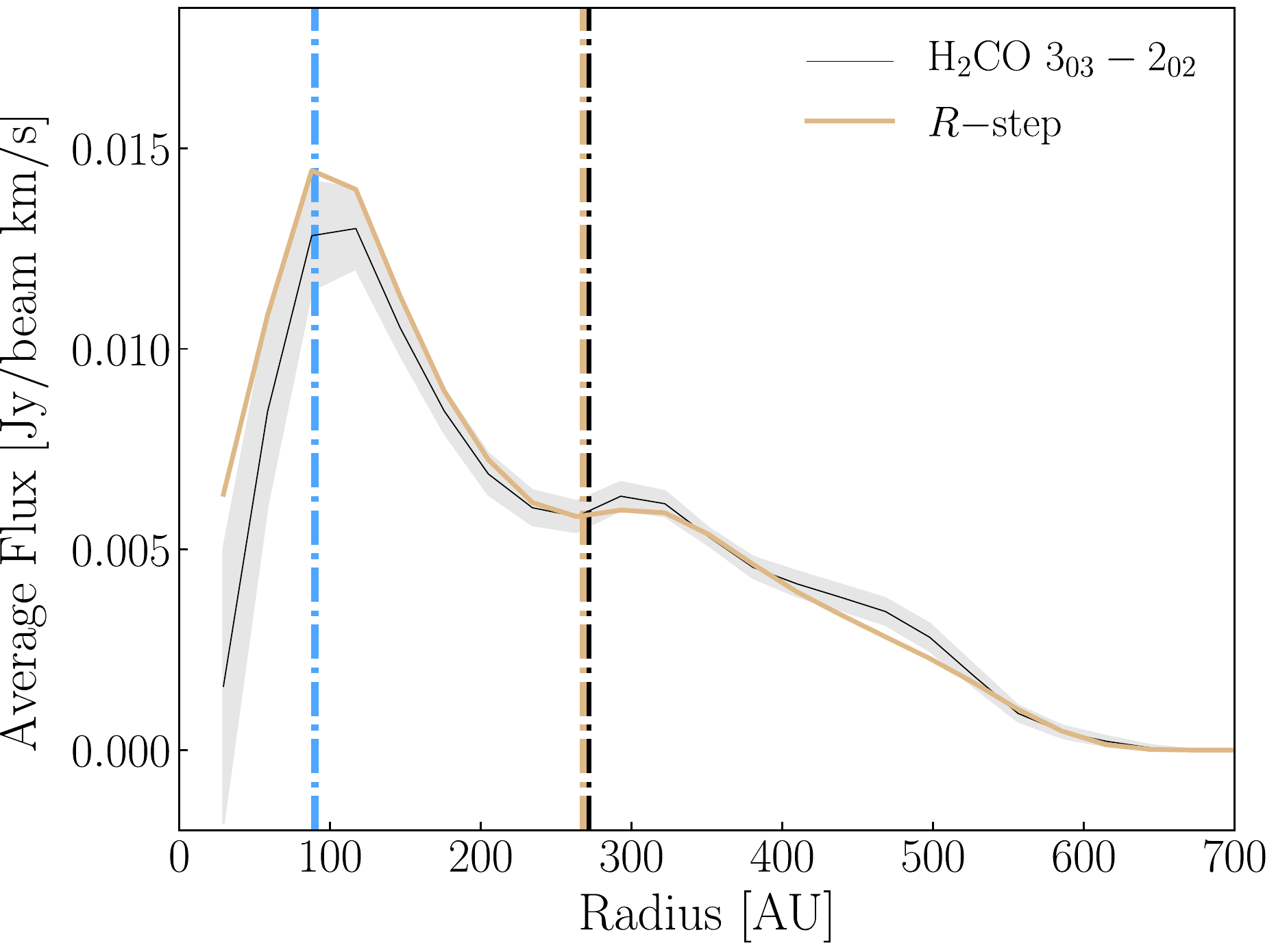} 
 \hspace{0.5cm}
 \centering
 \includegraphics[width=0.45\textwidth]{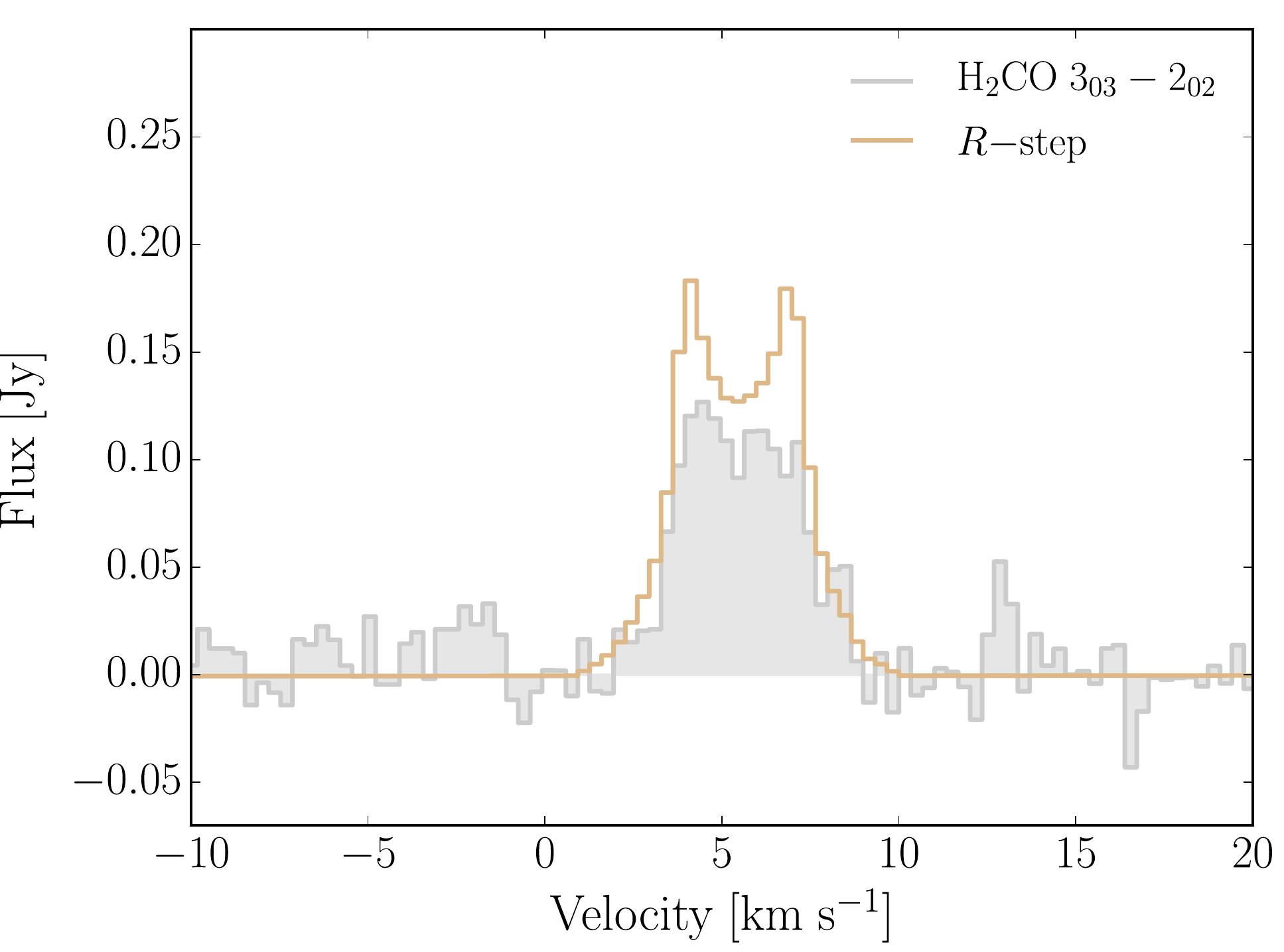} \\ 
 \raggedleft
 \hspace{-0.2cm}
 \includegraphics[width=0.435\textwidth,height=6cm]{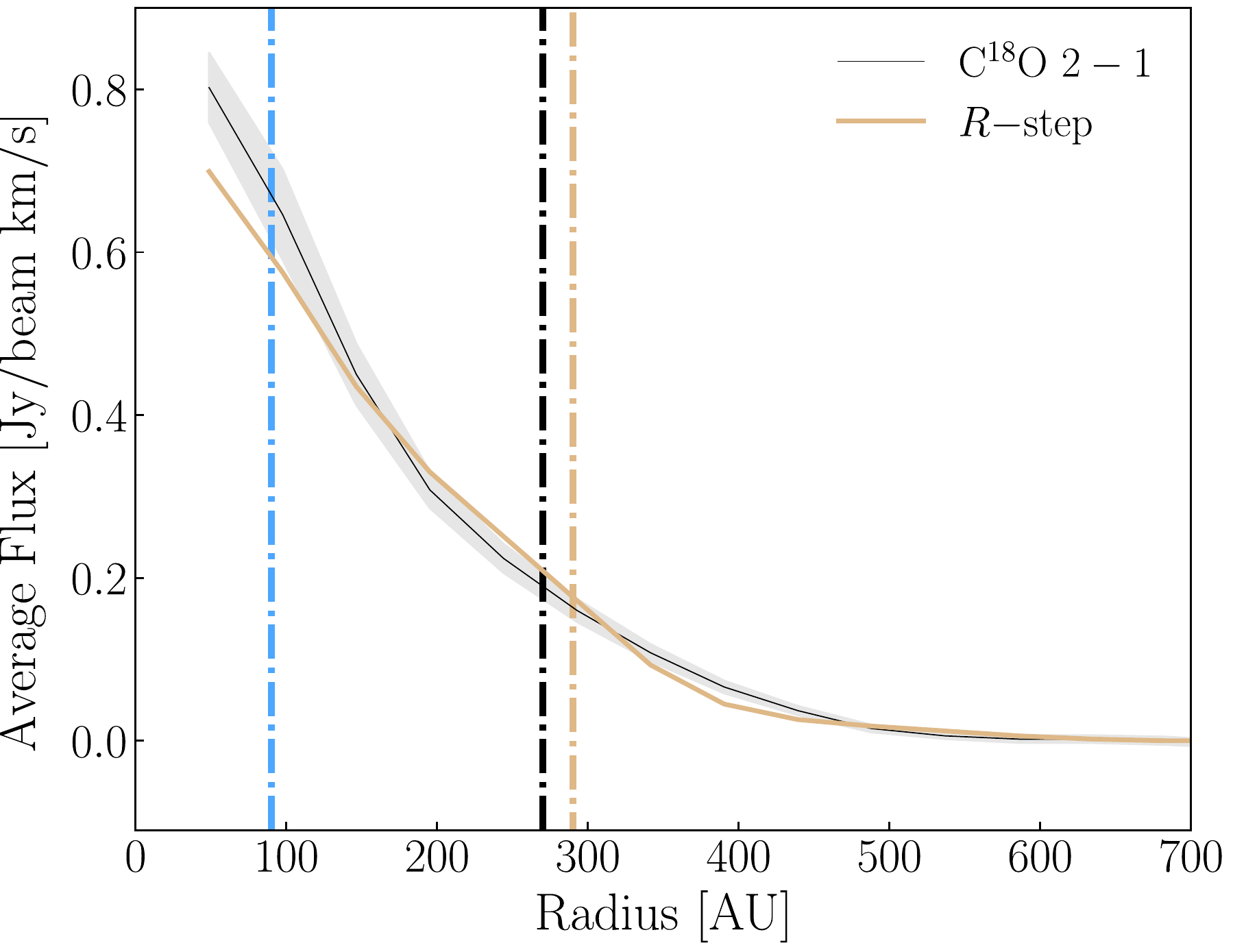} 
 \hspace{0.9cm}
 \centering
 \includegraphics[width=0.43\textwidth]{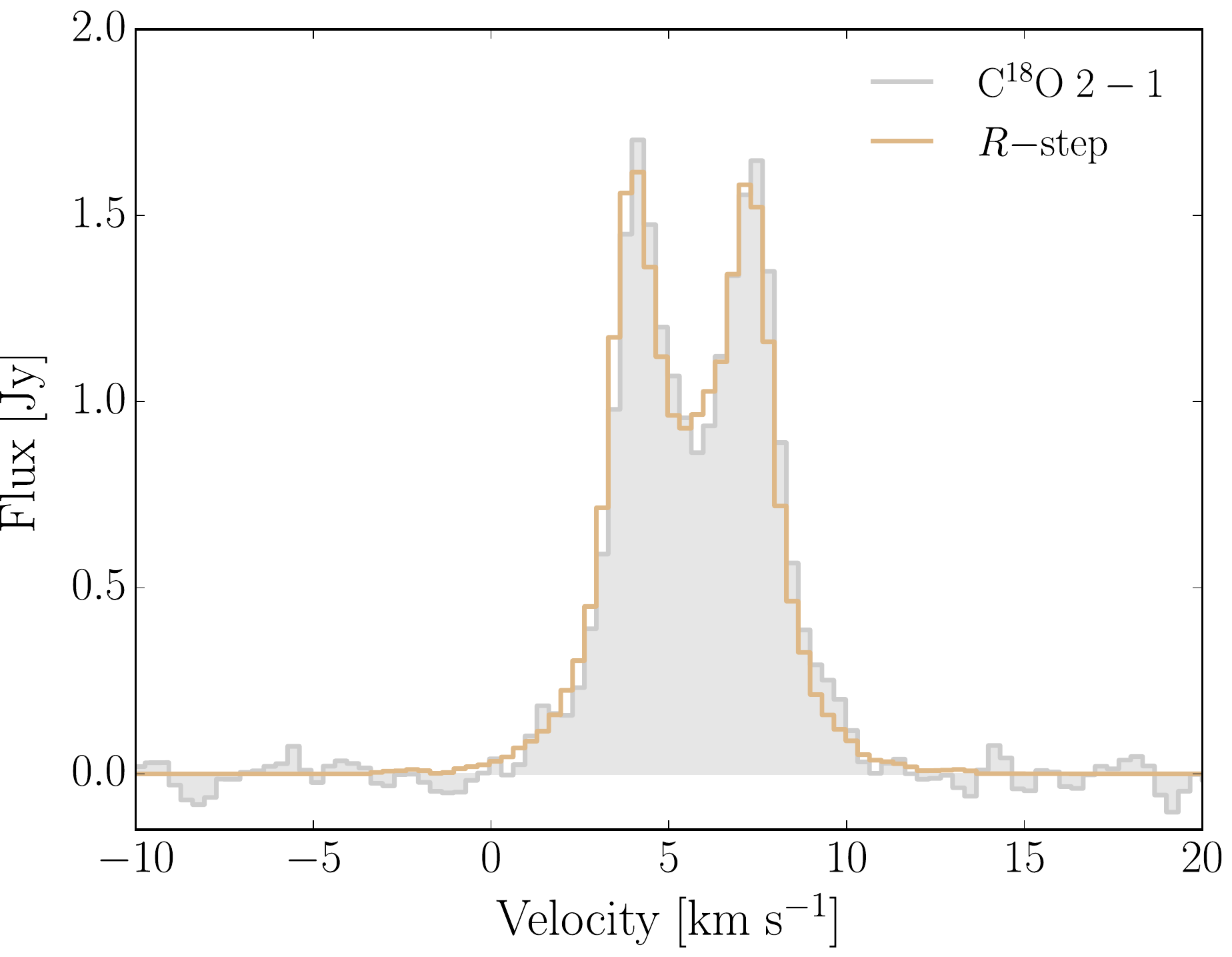} 
 \hspace{-0.6cm}
 \caption{Radial intensity and spectra of observed H$_2$CO 3$_{03}$--2$_{02}$ and C$^{18}$O 2--1 
 versus the best-fit models.
 {\it (Left)} Radial intensity curves from \change{azimuthally-averaged elliptical annuli projected to $i$=44$^{\circ}$, P.A.=133$^{\circ}$.}
 HD 163296 data is show in black, best-fits for H$_2$CO and C$^{18}$O are in gold.
 The vertical dashed lines indicate the CO snow line (blue dash) from \citet{Qi2015}, the 5$\sigma$ outer radius
 of the 1.3 mm grains (black dash), and the \change{change-over radii, $R_{\rm c}$, for the best-fit
 radial step-abundance models} (gold dash). \change{H$_2$CO profiles are taken from integrated intensity maps after applying a Keplerian mask.}
 {\it (Right)} Disk-integrated spectra. HD 163296 data is shown in filled gray.
 \change{H$_2$CO spectra are Hanning smoothed to 0.336 km $^{-1}$ channels.}}
 \label{fig:bestfitmod}
\end{figure*}

To estimate the absolute fractional abundances \change{relative to H$_2$} in the inner and outer 
regions for H$_2$CO and C$^{18}$O, LIME was used to \change{vary the 
abundances} for the best-fit normalized scenarios. Abundance
ratios across the change-over boundaries, $R_{\rm c}$, were kept the same as the 
normalized models: $X_1/X_2 = 0.5$ for the H$_2$CO radial step-abundance model
and $X_1/X_2 = 10$ for the C$^{18}$O radial step-abundance model. 

H$_2$CO models had $R_{c}$ = 270 AU 
and $X_1$ = [1.0, 2.0, 3.0, 4.0, 5.0] $\times \ 10^{-12}$.
The best-fit fractional abundances were $X_1 =  4.0 \times 10^{-12}$ and
$X_2 = 8.0 \times 10^{-12}$. 
C$^{18}$O was found to have best-fit fractional abundances of
$X_1 = 5.0 \times 10^{-8}$ and $X_2 = 5.0 \times 10^{-9}$
\change{with $R_{c}$ = 290 AU. Radial intensity profiles for 
these best-fit models are shown in Figure~\ref{fig:bestfitmod}. Error estimates
based 3$\sigma$ error bars of the radial intensity profiles
put these abundances in the range $X_1 =  2\!-\!5 \times 10^{-12}$, 
$X_2 = 5\!-\!10 \times 10^{-12}$ for H$_2$CO and 
$X_1 = 4\!-\!12 \times 10^{-8}$ and $X_2 = 4\!-\!12 \times 10^{-9}$ for C$^{18}$O.}

\begin{figure*}[!htbp]
 \centering
 \includegraphics[width=\textwidth]{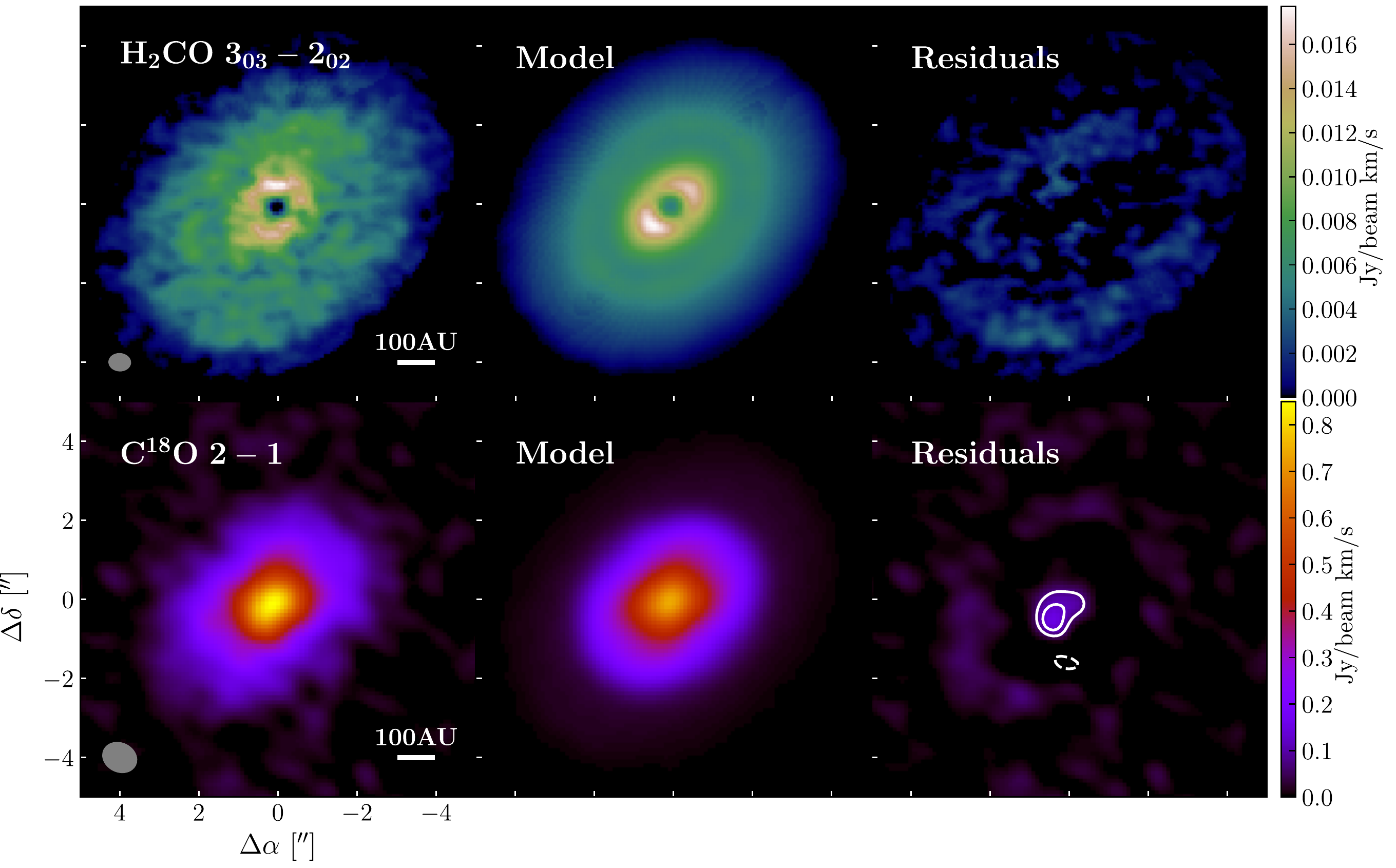}
 \caption{\change{Data, model, and residual integrated intensity maps.
 H$_2$CO 3$_{03}$--2$_{02}$ data and model maps are created
 after applying a Keplerian mask to the image cube (see Section~\ref{sec:res_distr}).}
 {\it Left:} H$_2$CO 3$_{03}$--2$_{02}$ data integrated intensity map from 0.76 -- 10.84 km s$^{-1}$ and
 C$^{18}$O 2--1 data integrated intensity map from --0.88 -- 12.48 km s$^{-1}$.
 Synthesized beam and AU scale are shown in the lower corners.
 {\it Center:} Integrated intensity map of best-fit model taken over the same velocity channels as the left figure.
 {\it Right:} Residual image with contours at 3$\sigma$ intervals. Dashed contours are negative, 
 solid contours are positive.}
 \label{fig:mom0modres}
\end{figure*}

Integrated intensity maps of the best-fit models were compared to integrated intensity
maps of observed H$_2$CO 3$_{03}$--2$_{02}$ and C$^{18}$O 2--1 data. Figure~\ref{fig:mom0modres}
shows the images and the residuals. The model and the 
data are in good agreement for both lines, though the best-fit C$^{18}$O
has residual emission above the 3$\sigma$ in the central part of the disk.
\change{The inner 50 AU are likely not well-described by our models, as
noted in Section~\ref{sec:res_mods}.}

The modeling efforts presented here show that the H$_2$CO abundance
is not uniform throughout the disk. Beyond $\sim$300 AU there is an increase
in the H$_2$CO abundance by a factor of two, as seen in the radial step-abundance
scenario. 
The H$_2$CO abundance of $X_1 = 2\!-\!5 \times 10^{-12}$, $X_2 = 4\!-\!10 \times 10^{-12}$
is \change{consistent to within a factor of a few} with the global abundance value of $1 \times 10^{-11}$
found in \citet{Qi2013}. The increased sensitivity and resolution of our data allow 
us to better constrain the H$_2$CO abundance in HD 163296 than previous studies.
C$^{18}$O is well-described by a model with a depletion of CO at 290$\pm$20 AU
and a depletion factor of ten. The C$^{18}$O inner abundance of $4\!-\!12 \times 10^{-8}$
corresponds to a $^{12}$CO abundance of $2.2\!-\!6.6 \times 10^{-5}$, assuming $^{12}$CO/C$^{18}$O = 550.
\citet{Qi2015} report similar numbers for the CO abundance,
but their depletion factor is lower by half \change{and occurs
at a radius of 90 AU.
We found that a radius of 90 AU and depletion factor of 5 for our radial step-abundance
models significantly overproduces the amount of C$^{18}$O beyond 300 AU
due to our different treatment of the vertical structure.}

\subsection{H$_2$CO excitation temperature}
\label{sec:res_extemp}

Line flux ratios H$_2$CO 3$_{03}$--2$_{02}$/H$_2$CO 3$_{22}$--2$_{21}$
and H$_2$CO 3$_{03}$--2$_{02}$/H$_2$CO 3$_{21}$--2$_{20}$
were used to constrain H$_2$CO excitation temperatures.
Table~\ref{tab:obs_par} provides the line fluxes. \change{We
calculated the rotational temperature of the lines,
assuming a single rotational temperature, following \citet{Qi2013}}

\begin{equation}
 \label{eq:trot}
 T_{\rm rot} = \frac{E_1 - E_0}{\mathrm{ln}((\nu_1 S \mu_1^2 \int T_0 d\nu) / (\nu_0 S \mu_0^2 \int T_1 d\nu))},
\end{equation}

\noindent with the following definitions: $E_0$ and $E_1$ are the upper energy levels 
for the low and high H$_2$CO transitions, respectively; 
$\nu$ is the line frequency; $S\mu^2$ is the temperature-independent transition 
strength and dipole moment; and $\int T d\nu$ is the integrated line intensity.
Line intensity in the Rayleigh-Jeans limit was calculated from the 
line flux with the following expression

\begin{equation}
 \label{eq:jytok}
 T_{\rm B} = \frac{c^2}{2 k \nu^2} \frac{F_{\nu}}{(a \times b)} \left(3600 \frac{\rm arcsec}{\rm deg}\right)^2 \left(\frac{180}{\pi} \frac{\rm deg}{\rm sr}\right)^2 \left(\frac{1}{10^{26} {\rm \ Jy}}\right),
\end{equation}

\noindent \change{where $F_{\nu}$ is the line flux in Jy, $T_{\rm B}$ is the 
line intensity in Kelvins, $\nu$ is the line frequency in Hz, $k$ is the Boltzmann constant,
$c$ is the speed of light, and $a$ and $b$ are the semi-major and semi-minor axes of the beam in arcsec.}

The emitting regions of all three lines
are expected to be similar, especially if the H$_2$CO reservoir is primarily
locked up in icy grains. LTE is a fair assumption for calculating rotational
temperatures, as the gas density near the midplane is high in disks 
\citep[$\sim$10$^9$ cm$^{-3}$;][]{Walsh2014} and the critical densities of the observed
transitions at 20 K are 1--3$\times$10$^6$ \citep{Wiesenfeld2013}.
In the case of LTE, the derived rotational temperature is equal to the
kinetic temperature of the gas.
$E$ and $S\mu^2$ are \change{taken from the CDMS \citep{Muller2005}, as 
reported on the Splatelogue\footnote{\url{http://www.cv.nrao.edu/php/splat/}} database.}

The rotational temperatures of the H$_2$CO transitions are calculated based on
the line flux ratios of H$_2$CO 3$_{22}$--2$_{21}$/3$_{03}$--2$_{02}$ and 
H$_2$CO 3$_{21}$--2$_{20}$/3$_{03}$--2$_{02}$. 
\change{The matched-filter technique only gives lower limits to the 
H$_2$CO 3$_{22}$--2$_{21}$ and H$_2$CO 3$_{21}$--2$_{20}$ line flux, 
thus lower limits on the rotational temperature are $>$20.5 K and $>$19.5 K, respectively, 
while upper limits for the weak lines are $<$169 K and $<$326 K 
based on the integrated flux upper limits listed in Table~\ref{tab:obs_par}.}
\change{These lower limits} indicate that these transitions \change{can be} excited in regions
of the disk near the CO freeze-out temperature, supporting the hypothesis
that some of the H$_2$CO emission may originate from the cold molecular reservoir.
There could also be H$_2$CO emitting at a higher temperature that is not well 
described by our template filter.


\section{Discussion}
\label{sec:disc}

In this work, the radial step-abundance model suggests 
an enhancement in H$_2$CO abundance by a factor of a few 
beyond 270 AU. It is difficult to distinguish which
formation route is responsible for this modest increase in
abundance. 

\citet{Aikawa1999} estimated the radial column density 
and abundance profile of H$_2$CO formed in the gas phase
in a T Tauri minimum mass solar nebula (MMSN) disk model
extrapolated out to $R_{out}$ = 700 AU, with an order of magnitude
lower mass. They did not consider other mechanisms
for producing gas-phase H$_2$CO, such as desorption from icy grains.
They exclude an activation energy barrier 
for the CH$_3$ + O reaction. The initial abundance of atomic oxygen
may affect the inferred H$_2$CO abundances.
Their model has a mostly flat radial distribution, 
but is consistent with an
enhancement of H$_2$CO abundance by a factor of 
a few up to one order of magnitude in the outer regions 
beyond $\sim$300 AU. 

\citet{Walsh2014} created a series of increasingly complex 
T Tauri disk chemical evolution models that include  
grain-surface formation to estimate abundances of complex
organic molecules (COMs) throughout the disk. 
Beginning with freeze-out and thermal desorption only, they
go on to include non-thermal desorption, grain-surface chemistry, 
radiative reprocessing of ices, and reactive desorption in their full disk model.
The vertical distribution of H$_2$CO included a large 
gas-phase reservoir above the midplane reaching peak fractional abundances
relative to H$_2$ of $\sim$10$^{-8}$ and an ice reservoir
beyond 10 AU with peak fractional abundances of $\sim$10$^{-4}$. 
Beyond 50 AU, the radial column density of H$_2$CO in the \citet{Walsh2014} 
comprehensive disk model shows an increase by a factor of a few.

\change{From these two examples it is clear that a modest
outer disk enhancement of H$_2$CO cannot immediately reveal
whether gas-phase or grain-surface production is the 
dominant formation route. 
Full chemical modeling of H$_2$CO production is required.} 
In this section we discuss
possible explanations for the H$_2$CO enhancement
in the outer disk around HD 163296, its relation to the
CO snow line and the millimeter continuum, and the implications
for H$_2$CO formation.

\subsection{H$_2$CO and the CO snow line}
\label{sec:disc_h2co_cosnowline}

Previous SMA observations of H$_2$CO in the disk around HD 163296
showed ring-like formaldehyde emission outside the expected CO snow line
\citep{Qi2013}. The authors suggested that a scenario with only grain-surface formation could be
responsible for the observed distribution and the apparent 
lack of centrally peaked emission. \change{The lower spatial 
resolution and SNR per channel of the SMA observations would preferentially
place the H$_2$CO emitting region farther away from the central star
since the emission at smaller radii is spread out over more velocity 
channels due to the shear in the Keplerian disk, thus resulting in 
a false ring-like structure.} The ALMA results presented here
show that H$_2$CO is not present in a ring, but rather emission is seen
throughout most of the gaseous extent of the disk, with a central
depletion in the inner $\sim$50 AU.

\citet{Qi2015} presented new constraints on the CO snow line
in HD 163296 based on observations of C$^{18}$O and N$_2$H$^+$.
N$_2$H$^+$ is readily destroyed
by gas-phase CO, thus it is expected to be a reliable tracer
of CO depletion.
\change{By refitting the location and degree of CO depletion,
they found that a factor of 5 depletion in
column density at 85--90 AU improved their best-fit models 
to the visibility data. They interpret this radius as the 
location of the CO snow line, corresponding to a 
CO freeze-out temperature of 25 K. The coincident
of CO depletion and N$_2$H$^+$ emission inner radius supports the claim
that N$_2$H$^+$ traces regions of CO freeze-out. Recent results by
\citet{vantHoff2017} show that the N$_2$H$^+$ emission can 
peak from $\sim$5--50 AU beyond the location of the CO snow line and 
that careful chemical modeling is necessary to properly interpret
the location of CO freeze-out from N$_2$H$^+$ observations.}

\change{The data presented here show that H$_2$CO extends beyond 
the \citet{Qi2015} CO freeze-out radius, but with a peak at $\sim$90 AU
that coincides with the CO snow line.
\"{O}berg et al. (in press) presented H$_2$CO observations in the disk around
TW Hya and find that grain-surface formation of H$_2$CO begins at 
temperatures where CO starts to spend even a short time on the grains, meaning that H$_2$CO can 
be produced -- and the emission can peak -- just inside of the CO snow line.
Considering our $\sim$50 AU resolution, the peak seen at $\sim$90 AU
may be the beginning of grain surface formation of H$_2$CO, likely with
some contribution from the warmer, gas-phase formation pathway at the innermost radii.}

Recent analysis of ALMA Cycle 0 data for H$_2$CO in DM Tau explored
the relative contributions of gas-phase and grain-surface 
formation pathways \citep{Loomis2015}. 
Their chemical models required 
both formation via gas-phase reactions and hydrogenation of CO ice 
in the outer regions of the disk to reproduce the centrally peaked
and outer disk emission. Our simple parameterized models do not
include chemical processing, but the presence of H$_2$CO at radii 
beyond the expected CO snow line at 90 AU and where millimeter grains are present
indicates that grain-surface formation is a \change{partial} contributor to the
H$_2$CO reservoir in the disk around HD 163296. \change{Unlike the DM Tau results,
our data also suggests that there is an intrinsic link between the edge
of the millimeter continuum and the production of H$_2$CO in HD 163296.}


\subsection{The H$_2$CO inner hole}
\label{sec:disc_h2co_hole}

A sharp drop in H$_2$CO emission within 50 AU is evident in the integrated intensity
map (see Figure~\ref{fig:mom0modres}) and by the best-fit $R_{\rm in}$
of the H$_2$CO models. \change{Optically thick dust may be} responsible 
for the observed depletion, rather than a drop in H$_2$CO abundance. 
Photons emitted by H$_2$CO in the midplane of the disk can be absorbed 
by optically thick dust in the upper layers, causing the inner hole -- after
continuum subtraction -- that is seen in the H$_2$CO integrated intensity map.

\citet{Zhang2016} model the HD 163296 continuum visibilities using a parameterized
radial intensity distribution modulated by multiple sine waves.
Their best-fit model shows that there is an increase in millimeter-wavelength
intensity of $\sim$60\% in the innermost 50 AU of the disk, causing
the millimeter continuum to become optically thick in this region.
\citet{Zhang2016} produced simulated model images with a 0.035$\arcsec$ beam, 
which clearly shows the strong central continuum emission. 
At the spatial resolution presented in this work, 0.5$\arcsec$, the 
millimeter continuum appears smooth. \change{\citet{Isella2017} presented 0.2$\arcsec$
observations of the 1.3 millimeter continuum and three CO isotopologues. 
They show central depressions in the $^{13}$CO and C$^{18}$O maps,
and concluded that both the CO and the dust become optically thick 
within 50 AU, leading to large uncertainties in their surface densities.}


\change{\"{O}berg et al. (in press) observe a similar central H$_2$CO depression well
within the CO snow line in the disk around TW Hya. While they do 
not rule out dust opacity effects completely,
they prefer to explain it as a real drop in abundance, as the central
depression is not seen in higher frequency lines of H$_2$CO or CO isotopologues.
Observations of additional, high frequency H$_2$CO lines in the HD 163296 disk would
be needed to test this scenario. The $^{13}$CO and C$^{18}$O central holes 
seen by \citet{Isella2017} at high resolution already suggest dust opacity as an explanation 
for the central H$_2$CO depression in this disk.}


\subsection{H$_2$CO and the millimeter continuum edge}
\label{sec:disc_h2co_mmedge}
The millimeter grains in HD 163296 have decoupled from
the gas and drifted radially inward.
Millimeter emission in the outer disk is truncated at 270 AU while
the bulk of the gas, based on $^{12}$CO observations, extends to $\sim$550 AU
\citep{deGregorioMonsalvo2013}. The outer edge of 
millimeter emission corresponds to the 270$\pm$20 AU 
change-over radius for H$_2$CO enhancement
found by the best-fit radial step-abundance model. 
\change{Grain growth and radial drift in the outer regions of the disk 
can result in a decrease of small, micron-sized grains
beyond $\sim$270 AU. With less shielding 
from external and protostellar radiation,  
penetrating UV photons in the outer regions may 
cause an increase in the H$_2$CO photodesorption 
directly off icy grain surfaces \citep{Oberg2009,Huang2016}.
Increased UV radiation in a dust-depleted outer disk
could also lead to CO photodissociation in
the upper layers, opening a more efficient gas-phase route for H$_2$CO
where hydrocarbon radicals and atomic oxygen are readily available.}

\change{Dust evolution models for HD 163296 by Facchini, et al. (subm.) show that for
a low-turbulence environment, a temperature inversion can occur around 300 AU,
causing a second CO desorption front in the outer disk
\citep[also suggested qualitatively by][]{Cleeves2016}.
In that case, an increase of both C$^{18}$O and H$_2$CO abundance 
in the outer disk would be expected, but our models found a CO depletion.  
It may be that there are competing effects occurring in the outer disk for CO.
A temperature inversion and/or UV photodesorption beyond 300 AU can cause the release
of grain-surface H$_2$CO and a fraction of CO ice back into the gas 
phase near the midplane, but CO photodissociation 
in the upper layers may dominate the C$^{18}$O surface density profile so that we ultimately
see an outer disk depletion in C$^{18}$O, and an increase in H$_2$CO production.}



\section{Conclusions}
\label{sec:concl}

In this work, multiple detections \change{with ALMA} of H$_2$CO 3--2 in the
protoplanetary disk around HD 163296 were presented: 
one robust detection via imaging, H$_2$CO 3$_{03}$--2$_{02}$,
and two weaker detections via matched filter analysis, 
H$_2$CO 3$_{22}$--2$_{21}$ and H$_2$CO 3$_{21}$--2$_{20}$.
The distribution of H$_2$CO
relative to C$^{18}$O and the millimeter continuum 
was analyzed using various model
abundance profiles to test possible H$_2$CO formation
scenarios. The conclusions of this work are as follows:

\begin{itemize}
  \item H$_2$CO in HD 163296 is observed out to $\sim$550 AU, 
  equal to the full radial extent of the gas disk as observed with CO.
  \change{It does not have a ring-like morphology.}
  \item The kinetic temperature of the observed H$_2$CO gas has
  a lower limit of $>$20 K,
  thus emission from these lines can originate from 
  the cold molecular reservoir near the disk midplane.
  \item The best-fit radial step-abundance model
  to the H$_2$CO 3$_{03}$--2$_{02}$ data
  suggests that H$_2$CO has an inner radius $R_{in}$ = 50 AU, an outer disk abundance
  a factor of \change{two} higher than the inner disk ($X_1/X_2 = 0.5$),
  and a change-over radius of $R_{\rm c}$ = 270$\pm$20 AU. \change{There is a mechanism causing
  increased H$_2$CO production in the outer disk beyond the millimeter grains.
  One explanation is desorption of H$_2$CO from icy grains by thermal desorption 
  due to a temperature inversion or by UV photodesorption where CO is frozen out.
  Alternatively, photodissociation of CO in the outer disk may increase
  the efficiency of the CH$_3$ + O gas-phase route to form H$_2$CO.}
  \item \change{Based on the C$^{18}$O 2--1 models presented in this work, 
  two scenarios reproduce the the data well: step-abundance models with 
  abundance boundaries based on temperature and radius, respectively. 
  The best-fit models both indicate depleted CO in the outer disk 
  based on the recovery of the C$^{18}$O 2--1 surface density profile. 
  Both models have a CO depletion factor of 10 in the outer disk. 
  The depletion is likely a combination of CO freeze-out in the disk midplane and 
  photodissociation of CO in the disk upper layers 
  due to penetrating UV radiation.}
  \item \change{The best-fit abundance for the C$^{18}$O radial step-abundance model 
  was $X_1 = 4\!-\!12 \times 10^{-8}$, $X_2 = 4\!-\!12 \times 10^{-9}$. The 
  best-fit abundance for the H$_2$CO radial step-abundance model was $X_1 = 2\!-\!5 \times 10^{-12}$, 
  $X_2 = 4\!-\!10 \times 10^{-12}$. }

\end{itemize}

Further observations of HD 163296 can unambiguously determine the \change{dominant} formation
pathway of H$_2$CO in the disk. Constraining the ortho- to para- ratio of the 
two H$_2$CO isomers can distinguish between grain surface formation and 
gas-phase formation \citep{Guzman2011}. \change{The H$_2$CO o/p ratio is 
expected to be less than three for grain-surface formation
\citep{Dulieu2011, Fillion2012}. Observations
of co-spatial H$_2$CO and CH$_3$OH would also constrain the contributions of 
gas- and solid-phase H$_2$CO, as CH$_3$OH forms similarly 
via hydrogenation of CO ices and has no known gas-phase formation route.} 

\begin{acknowledgements}
      The authors thank the referee for insightful comments and 
      constructive suggestions.
      M.T.C. thanks S. Facchini and G.S. Mathews for useful discussion on
      dust evolution and CO chemistry in HD 163296.
      The authors acknowledge support by Allegro, the European
      ALMA Regional Center node in The Netherlands, and expert
      advice from Luke Maud in particular.
      This paper makes use of the following ALMA data: ADS/JAO.ALMA\#
      2013.1.01268.S and 2011.1.00010.SV.
      ALMA is a partnership of ESO (representing its member states), 
      NSF (USA) and NINS (Japan), together with NRC (Canada), NSC and 
      ASIAA (Taiwan), and KASI (Republic of Korea), in cooperation with 
      the Republic of Chile. The Joint ALMA Observatory is operated by 
      ESO, AUI/NRAO and NAOJ.
\end{acknowledgements}


\bibliographystyle{aa}
\bibliography{carney_phd_paper1}

\begin{appendix}

  \section{Channel Maps}
  \label{app:A}
    
  \begin{figure*}[!htbp]
    \centering
    \includegraphics[height=10cm]{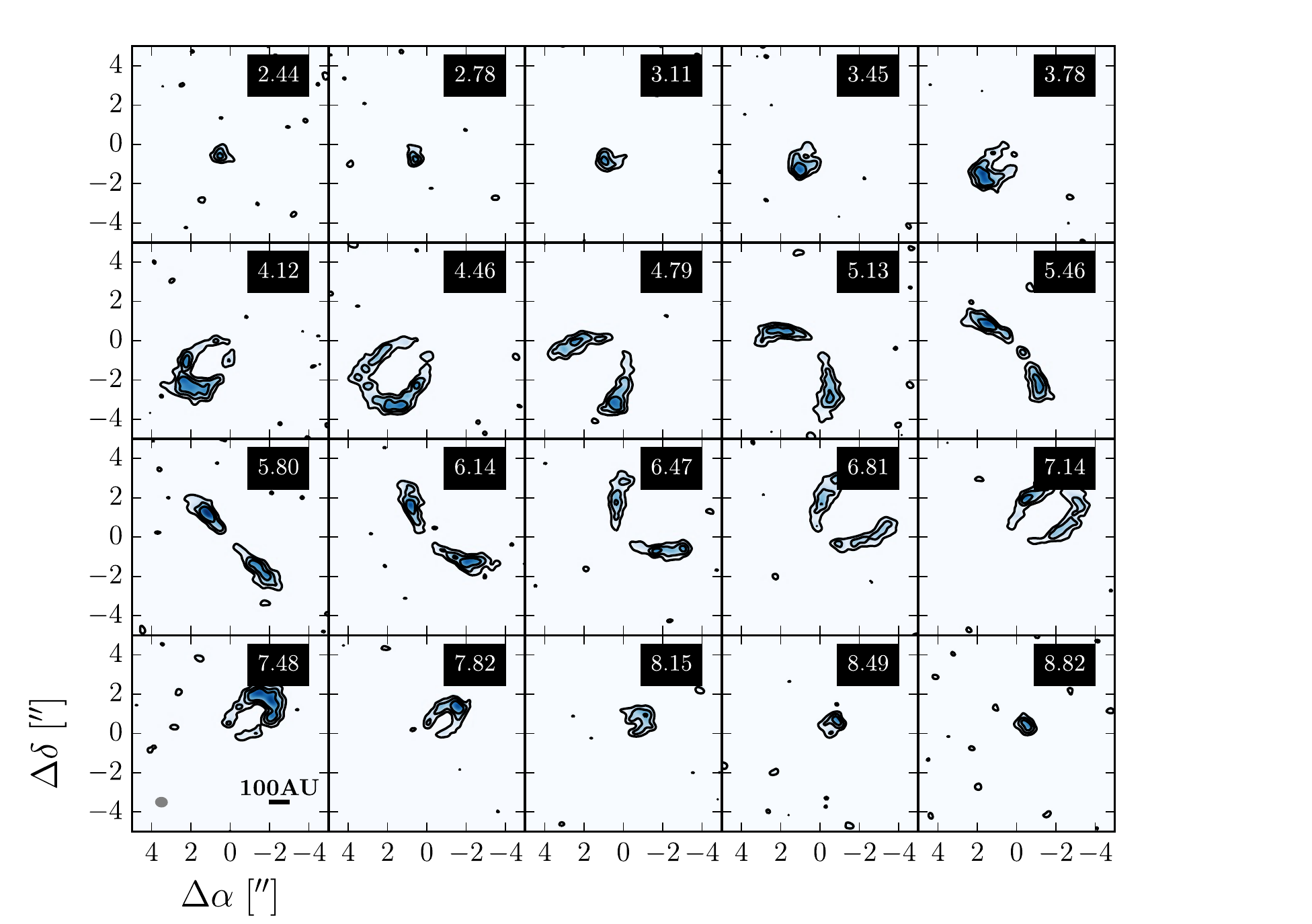}
    \caption{H$_2$CO 3$_{03}$--2$_{02}$ data channel maps, Hanning smoothed to 0.336 km s$^{-1}$ channels. 
    Black contours mark 1.5 $\times$ $10^{-3}$
    (1$\sigma$) $\times$ [3, 6, 9] Jy beam$^{-1}$.
    Synthesized beam and AU scale are shown in the lower left panel.}
  \end{figure*}

  \begin{figure*}[!htbp]
    \centering
    \includegraphics[height=10cm]{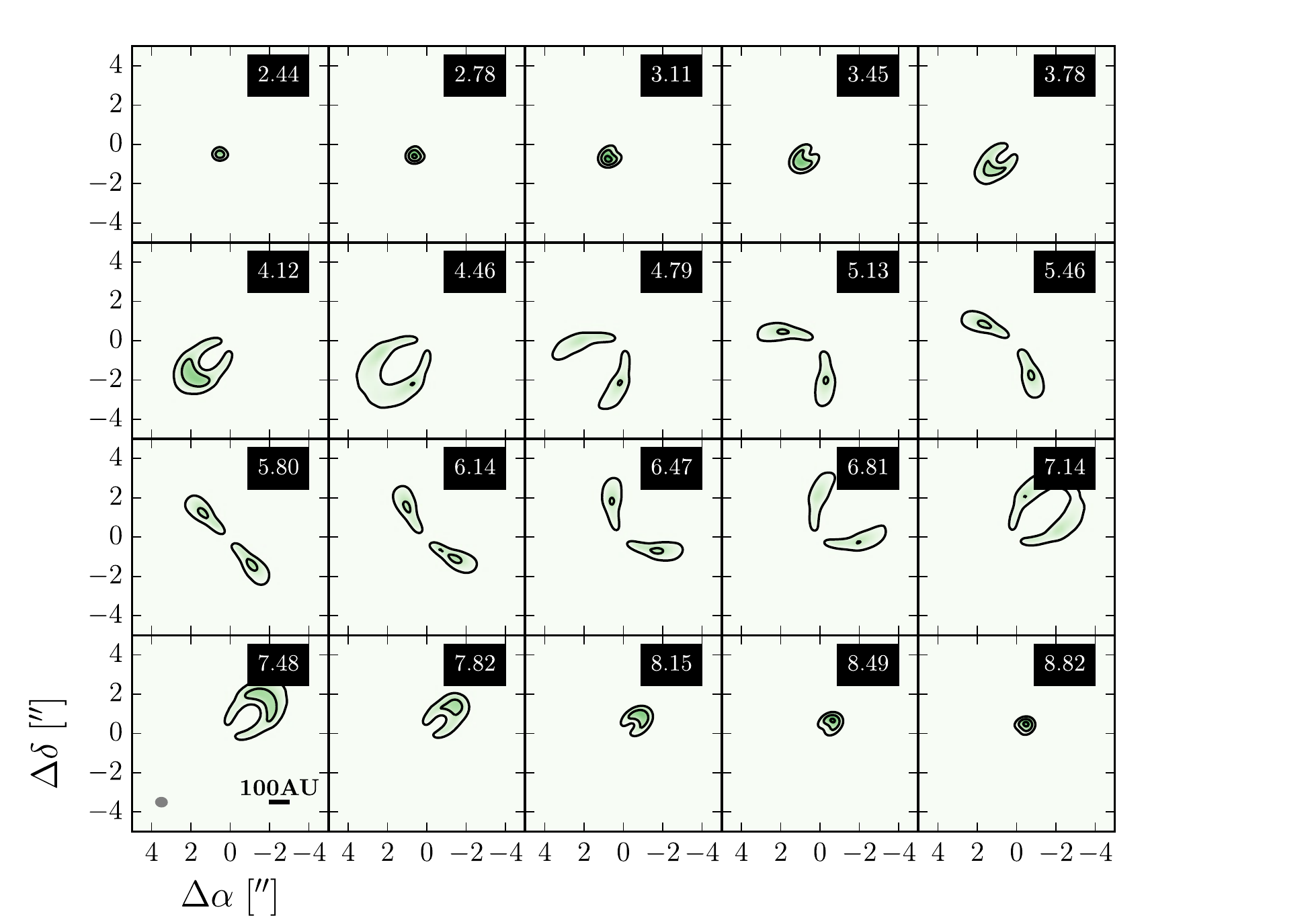}
    \caption{H$_2$CO 3$_{03}$--2$_{02}$ best-fit model channel maps, Hanning smoothed to 0.336 km s$^{-1}$ channels. 
    Black contours mark 1.5 $\times$ $10^{-3}$
    (1$\sigma$) $\times$ [3, 6, 9] Jy beam$^{-1}$.
    Synthesized beam and AU scale are shown in the lower left panel.}
  \end{figure*}
  
  \begin{figure*}[!htbp]
    \centering
    \includegraphics[height=10cm]{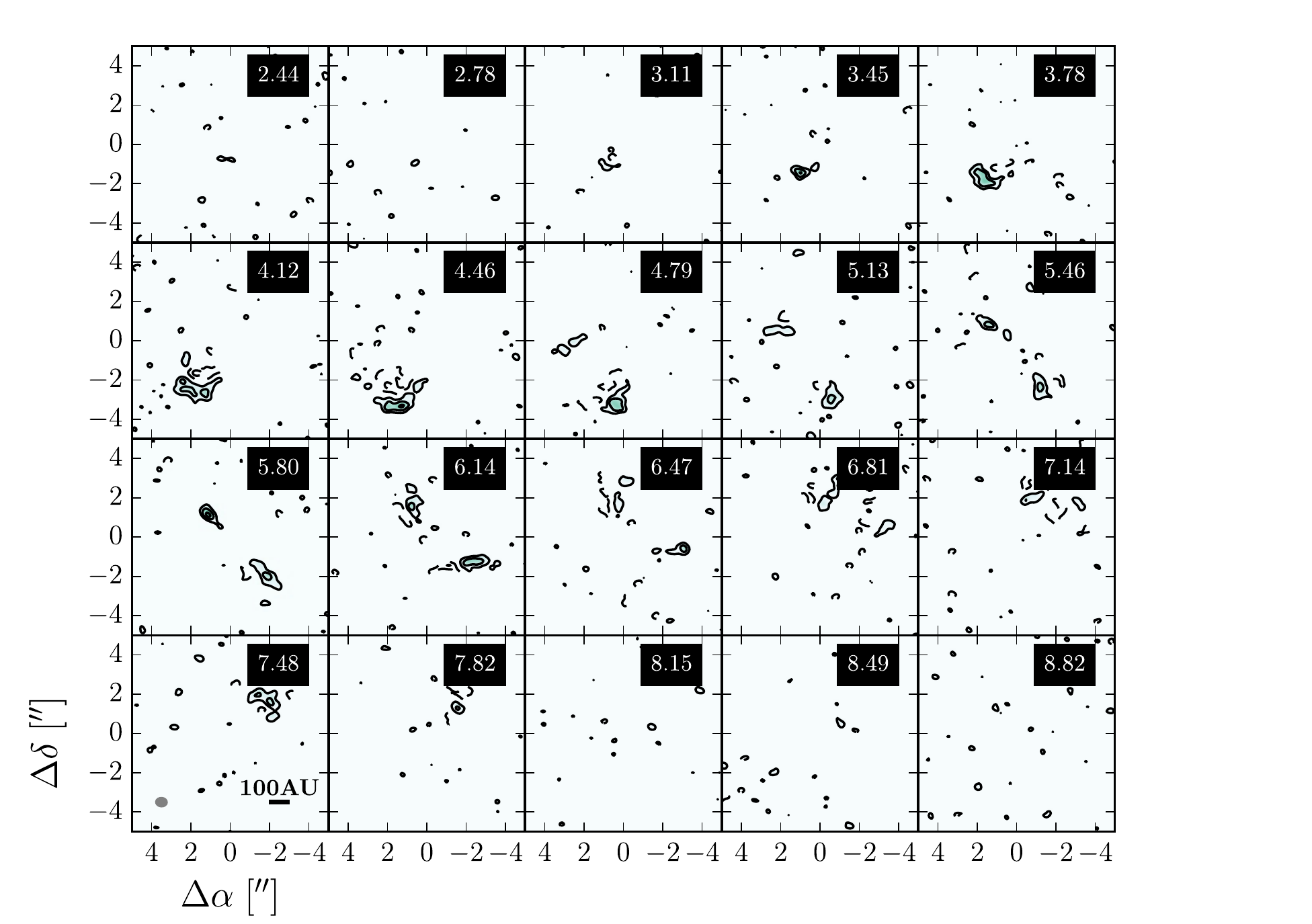}
    \caption{H$_2$CO 3$_{03}$--2$_{02}$ residual (data--model) channel maps, Hanning smoothed to 0.336 km s$^{-1}$ channels. 
    Black contours mark 1.5 $\times$ $10^{-3}$
    (1$\sigma$) $\times$ [3, 6, 9] Jy beam$^{-1}$. Dashed contours are negative at the
    same intervals. Synthesized beam and AU scale are shown in the lower left panel.}
  \end{figure*}
  
  \begin{figure*}[!htbp]
    \centering
    \includegraphics[height=10cm]{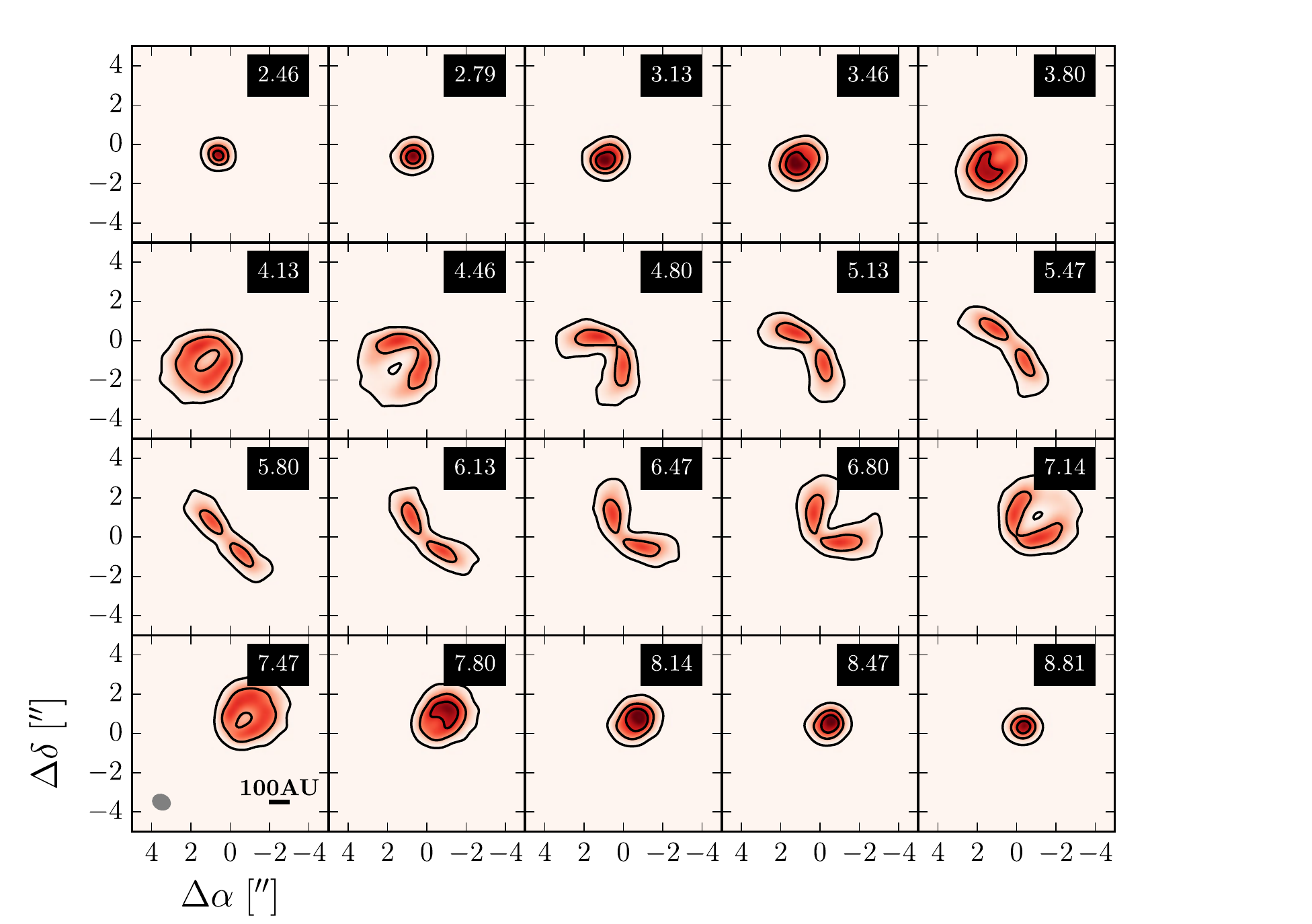}
    \caption{C$^{18}$O 2--1 data channel maps. Black contours mark 4.2 $\times$ $10^{-3}$
    (1$\sigma$) $\times$ [5, 25, 45] Jy beam$^{-1}$.
    Synthesized beam and AU scale are shown in the lower left panel.}
  \end{figure*}

  \begin{figure*}[!htbp]
    \centering
    \includegraphics[height=10cm]{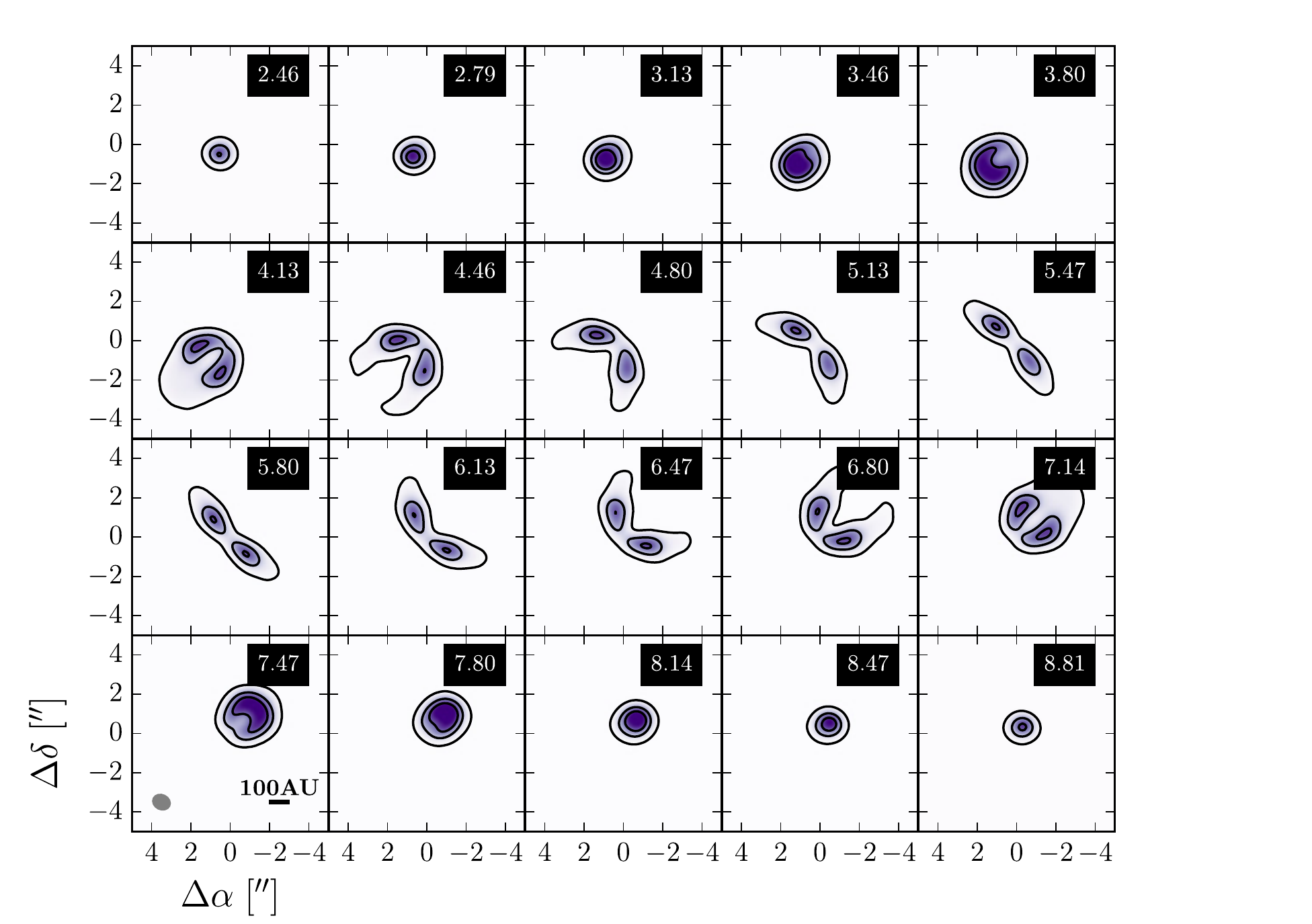}
    \caption{C$^{18}$O 2--1 best-fit model channel maps. Black contours mark 4.2 $\times$ $10^{-3}$
    (1$\sigma$) $\times$ [5, 25, 45] Jy beam$^{-1}$.
    Synthesized beam and AU scale are shown in the lower left panel.}
  \end{figure*}
  
  \begin{figure*}[!htbp]
    \centering
    \includegraphics[height=10cm]{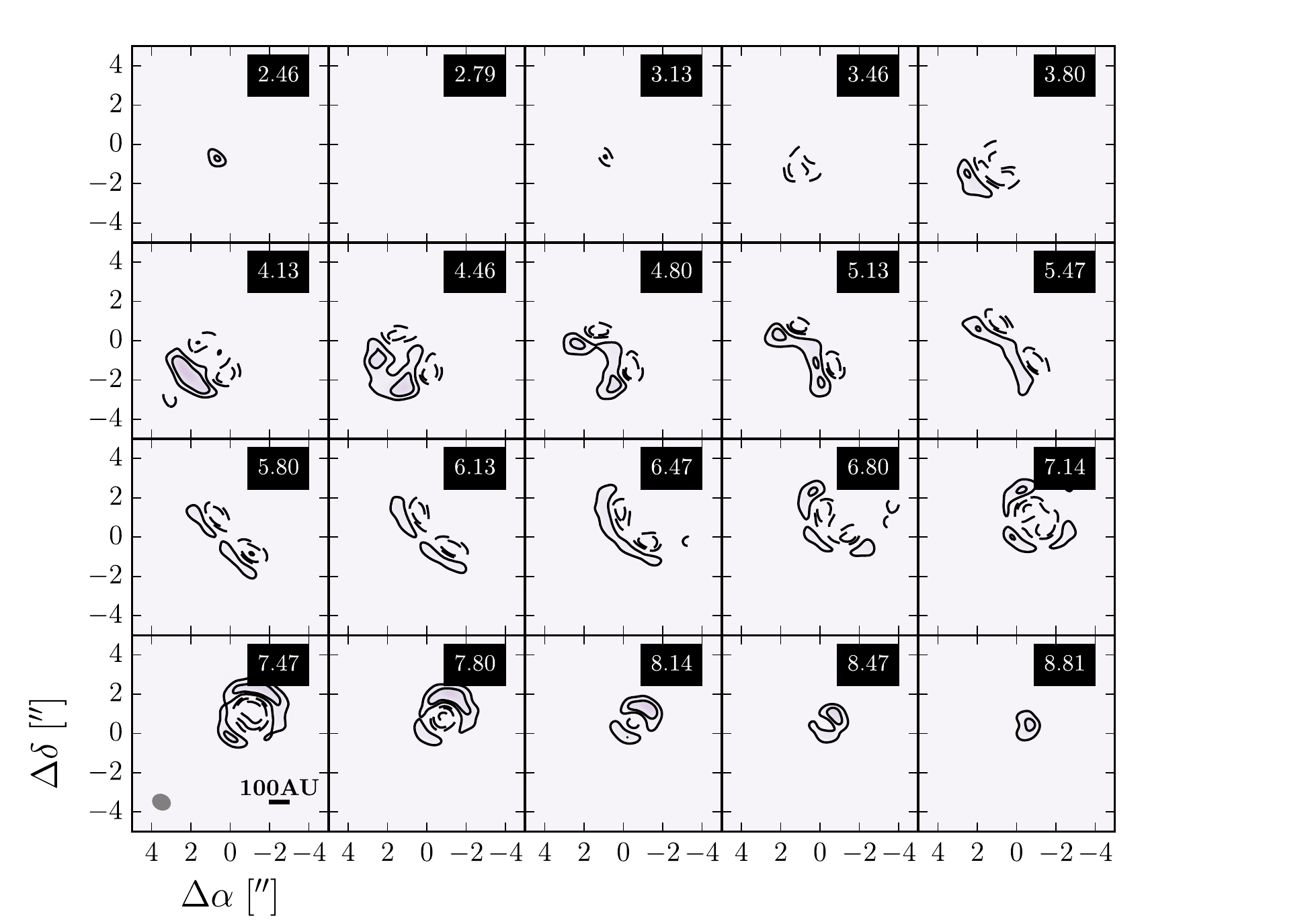}
    \caption{C$^{18}$O 2--1 residual (data--model) channel maps. Black contours mark 4.2 $\times$ $10^{-3}$
    (1$\sigma$) $\times$ [5, 10, 20] Jy beam$^{-1}$. Dashed contours are negative at the
    same intervals. Synthesized beam and AU scale are shown in the lower left panel.}
  \end{figure*}

\end{appendix}

\end{document}